\documentclass[num-refs]{wiley-article}

\usepackage{siunitx}
\usepackage{indentfirst}
\usepackage{amsmath}
\usepackage{caption}
\usepackage{subcaption}
\usepackage{parskip}
\usepackage{float}
\usepackage{multicol}
\usepackage{pdfpages}
\usepackage{threeparttable}
\usepackage{graphicx}
\usepackage{booktabs}
\usepackage{amssymb}
\usepackage{amsfonts} % for access to mathematical fonts
\usepackage{stmaryrd}
\usepackage{url}
\usepackage{booktabs}
\usepackage{multirow}
\usepackage{tabularx}
\usepackage[section]{placeins}
\usepackage{array}         % Custom column formatting
\usepackage[table]{xcolor} % For highlighting cells
\usepackage{colortbl}
\usepackage{ulem}
\definecolor{darkgreen}{rgb}{0.0, 0.5, 0.0}
\papertype{Original Article}
\paperfield{Journal Section}

\title{An Atlas of Extreme Properties in Cubic Symmetric Metamaterials}
\abbrevs{CNN, Convolutional Neural Network}

\author[1]{Sahar Choukir}
\author[1]{Nirosh Manohara}
\author[1,2]{\\Chandra Veer Singh}

\contrib[\authfn{1}]{Equally contributing authors.}

\affil[1]{Mechanical and Industrial Engineering, University of Toronto, Toronto, Ontario, M5S 3G8, Canada}
\affil[2]{Materials Science and Engineering, University of Toronto, Toronto, Ontario, M5S 3E4, Canada}

\corraddress{Chandra Veer Singh, University of Toronto, Toronto, Ontario, M5S 3E4, Canada}
\corremail{chandraveer.singh@utoronto.ca}

\begin{document}

\begin{frontmatter}
\maketitle

\begin{abstract}
Current research on three-dimensional metamaterial has largely focused on conventional strut, plate, and shell-based lattice designs. Although these designs offer several advantages, they possess inherent limitations that can restrict their performance in certain applications, motivating the exploration of alternative structural topologies. Here, we present a large-scale, symmetry guided framework for the generation and analysis of architected metamaterials based on all 36 cubic space groups. Using a voxel-based representation, we construct a database of approximately 1.95 million periodic unit cells spanning a broad range of relative densities and topological complexity. This dataset reveals a rich elastic property landscape shaped by crystallographic symmetry, including rare pentamode designs with high bulk to shear ratios such as $K/G\approx166$, isotropic-auxetic architectures with Poisson's ratio as low as $\nu=-0.76$, and structures achieving up to 93\% of the Hashin–Shtrikman upper bound on stiffness. Complementing the dataset, we develop a three-dimensional convolutional neural network surrogate model trained and evaluated on the full database to predict strain-energy density values under uniaxial, shear, and hydrostatic loading. Together, this work establishes a comprehensive atlas of cubic symmetric metamaterials and provides a pre-trained model for the accelerated discovery and design of 3D architected materials with extreme mechanical properties.

% Please include a maximum of seven keywords
\end{abstract}

\end{frontmatter}

\section{Introduction}

Mechanical metamaterials are architected materials whose effective properties emerge primarily from their underlying geometric structure rather than their chemical composition, enabling functionalities that extend far beyond those attainable in conventional bulk materials \cite{yu2018mechanical, barchiesi2019mechanical}. Through deliberate microstructural design, metamaterials have been shown to exhibit unconventional responses such as negative Poisson’s ratios \cite{Wilt2020} and exceptionally high stiffness-to-weight ratios \cite{zheng2014ultralight}. These capabilities unlock transformative opportunities across a wide range of engineering applications, including biomedical systems \cite{sun2024auxetic} and aerospace structures \cite{AirbusBionic2016}%\cite{du2020laser}
, where performance is often tightly constrained by weight and stiffness.

To date, much of metamaterial research has concentrated on a relatively narrow set of architectural families, most notably strut-~\cite{BastekJanHendrik2022Itsm}, plate-~\cite{MeyerPaulP2024NpRn}, and shell-based~\cite{PatelDarshil2023Dlid} networks. Although these designs have enabled a wide range of unconventional and desirable mechanical responses, they also exhibit inherent limitations, including stress concentrations at junctions and sensitivity to manufacturing imperfections and defects \cite{latture2018effects, liu2024ultrastiff, dastani2023effect}. To broaden the accessible design space, recent studies have explored alternative lattice classes, such as spinodoid metamaterials \cite{kumar2020inverse} and hybrid architectures that combine features from strut-, plate- and shell-lattices \cite{liu2024ultrastiff}. Crystallographic symmetry has emerged as a particularly powerful framework for systematic expansion of this design space. By enforcing global geometric constraints, symmetry operations directly govern load transfer pathways and deformation mechanisms, thereby shaping the effective mechanical response. Prior work has demonstrated the promise of symmetry-guided generation strategies. However, these efforts have remained limited in scope, focusing primarily on 2D lattices \cite{Mao2020} or 3D truss-based networks \cite{lumpe2021exploring}. As a result, the broader class of three-dimensional architectures featuring continuous surfaces, complex void morphologies, and hybrid plate-like geometries remains largely unexplored.

Traditional approaches to evaluate and screen such architectures rely heavily on Finite Element Method (FEM) simulations \cite{mizzi2018analytical, findeisen2017characteristics, zhao2015predicting, hsieh2019mechanical}, which can be both computationally intensive and time-consuming. This limitation has prompted a turn toward machine learning (ML) methods for accelerated characterization and design \cite{qi2019prediction, nath2024application}. In particular, deep learning techniques have shown promise in predicting effective material properties \cite{WiltJacksonK2020AAMD, JIANG2023100485} and guiding inverse design strategies to meet application-specific targets \cite{Abu2023, PahlavaniHelda2024DLfS, zheng2023unifying}.

Thus, this work introduces a systematic framework for constructing a large and diverse dataset of three dimensional metamaterial lattices derived from cubic crystal symmetries. To complement this novel dataset, we have developed a three-dimensional Convolutional Neural Network (CNN) model capable of accurately predicting the stiffness of these unique lattice structures from their voxelized representations. The distinctive aspect of this research is the explicit integration of crystallographic symmetry, enabling a rigorous investigation of how specific symmetry operations and Bravais lattices influence the emergence of extreme mechanical responses. In addition, the predictive 3D-CNN model can be utilized as a preliminary tool to FEM to screen for unique topologies with desired characteristics. This departure from conventional approaches aims to contribute to a deeper understanding and refine the predictive capabilities of models in the realm of metamaterial research, ultimately facilitating the optimization and discovery of high performance architectures for specific applications.

\section{Results}
\subsection{The Elastic Property Landscape}\label{subsec:prop-space}

The dataset encompasses 1.95 million voxelized unit cells systematically generated through cubic crystal symmetry operations spanning space groups 195–230, covering relative densities from 0.05 to 0.50 uniformly, yielding roughly 50k structures per symmetry on average. This broad coverage ensures that the dataset isn’t biased toward just a few well-known lattices; instead, it explores designs stemming from different symmetry operations (rotations, mirror planes, inversions) that were rarely considered in lattice literature. Notably, several previous studies have recognized the value of symmetry-guided design on superior mechanical properties but those were limited to strut-based graphs \cite{lumpe2021exploring} or two-dimentional lattices \cite{Mao2020}. In contrast, the present work’s voxel-based method can produce continuous surfaces and complex void shapes, not just truss networks. This allows discovery of plate- or shell-like features that strut-only approaches might miss. To the best of our knowledge, this vast library represents the most comprehensive exploration to date of how crystallographic symmetry governs mechanical response in 3D architected materials.

% The sheer scale of $\sim$ 1.95 million unique 3D unit cells is unprecedented and surpasses existing datasets. Thus, this vast library represents the most comprehensive exploration to date of how crystallographic symmetry governs mechanical response in three-dimensional architected materials. 

% As shown in Figure \ref{fig:property_overview}a, the distribution of structures is nearly uniform across symmetry classes, with space group 207 contributing the largest number of samples (87,325) and space group 230 the smallest (18,779), yielding an average of $\approx$ 54000 structures per group.

For each unit cell, we compute the Young’s modulus in all directions of the structure and denote the maximum and minimum values as $\overline{E}_{\text{max}}$ and $\overline{E}_{\text{min}}$, respectively. The mean stiffness is defined as the average of these two values, $\overline{E}_{\mathrm{mean}} = \tfrac{\overline{E}_{\text{max}} + \overline{E}_{\text{min}}}{2}$, while the variation in stiffness is expressed as $\Delta\overline{E} = \tfrac{\overline{E}_{\text{max}} - \overline{E}_{\text{min}}}{2}$. Anisotropy is quantified as $\Omega = \tfrac{\Delta\overline{E}}{\overline{E}_{\mathrm{mean}}}$, with structures satisfying $\Omega \le 0.05$ classified as isotropic. Mechanical optimality is defined relative to the Hashin-Shtrikman upper bound, with structures achieving $\tfrac{\overline{E}_{\mathrm{mean}}}{{E}_{\mathrm{HSU}}} \ge 0.9$ considered as optimal.  We further identify auxetic architectures with Poisson’s ratio $\nu \le 0$, thereby including the transition between non-auxetic and auxetic behavior. This definition mitigates sensitivity to numerical uncertainty in homogenized FEM predictions, though we note that strictly auxetic responses correspond to $\nu < 0$. The classification thresholds adopted here are consistent with previous benchmarks in metamaterial design \cite{Mao2020, meier2024obtaining}, establishing a unified framework for mapping and discovering architected metamaterials across the cubic-symmetry landscape.

\begin{figure}[!h]
\centering
\includegraphics[width=0.9\textwidth]{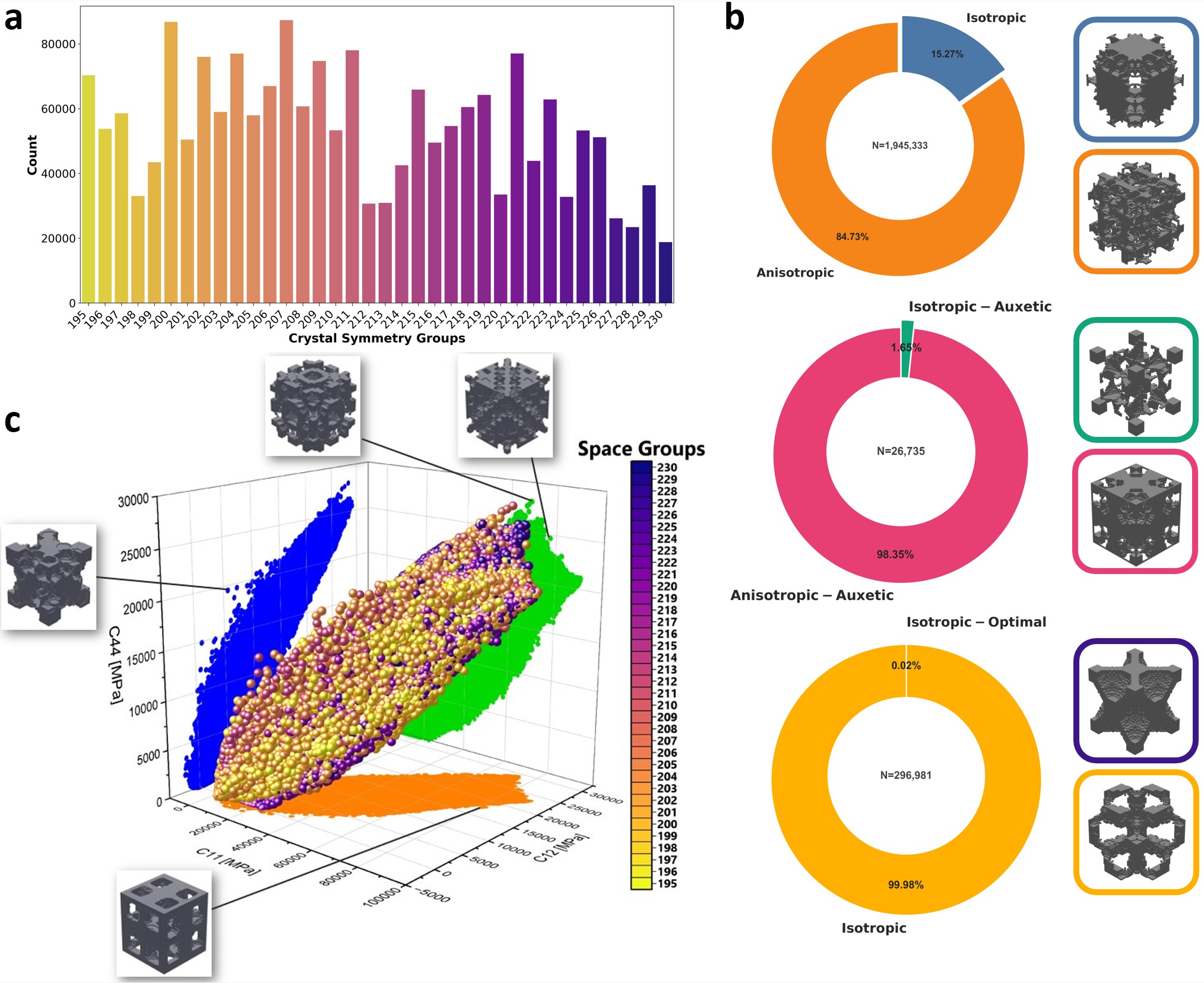}
\caption{Dataset composition and elastic-property landscape. \textbf{(a)} Frequency of samples per crystal symmetry group across the full dataset (\(N=1{,}945{,}333\)). \textbf{(b)} Structure statistics, displaying the proportion of anisotropic, isotropic, auxetic and optimal samples within the dataset, with sample unit cells belonging to those groups displayed alongside. \textbf{(c)} Distribution of stiffness in \((C_{11}, C_{12}, C_{44})\) space, colored by space group, with representative unit cell topologies highlighted in the property manifold.}
\label{fig:property_overview}
\end{figure}

Figure \ref{fig:property_overview}b highlights these isotropic, auxetic, and near-optimal subsets within the broader design space explored in this study. Figure \ref{fig:property_overview}c presents the distributions of elastic constants $C_{11}$, $C_{12}$, and $C_{44}$ across all space groups and randomly selected representative topologies, with most structures exhibiting modest stiffness. Only 15.27\% of the unit cells in the dataset are isotropic, for a total of 296,981, and an even smaller subset of 66 structures achieve isotropy while approaching the Hashin-Shtrikman upper bounds. Furthermore, the dataset contains 26,735 auxetic structures, of which 442 are isotropic. The relatively small fraction of isotropic-optimal and isotropic-auxetic architectures observed in the dataset underscores the inherent rarity of such designs, a trend that is consistent with their limited representation in the existing metamaterials literature.

\FloatBarrier

\FloatBarrier

\subsection{An Atlas of Extremes in Mechanical Metamaterials}

Here we identify several distinct families of structures that exhibit extreme and desirable mechanical responses, including isotropic-auxetic, isotropic-optimal, pentamode, and highly anisotropic architectures, as shown in Figure \ref{fig:extreme}. $E_{norm}=\frac{E}{E_{s}\overline{\rho}},  G_{norm}=\frac{G}{G_{s}\overline{\rho}}$ and $K_{norm}=\frac{K}{K_{s}\overline{\rho}}$ denote the Young’s modulus in the $\langle 100 \rangle$ directions, shear modulus, and bulk modulus normalized by the corresponding modulus of the parent material and the relative density of the architecture. In our analysis, for subsets that contain an abundance of geometries we select 200 structures that achieve the highest overall stiffness to plot, quantified by the average of these normalized moduli $Max (E_{norm} + G_{norm} + K_{norm}) / 3$, as indicated in the legend of Figure \ref{fig:extreme}. 

%\begin{equation} 
%\label{eq:E_norm}
%E_{norm}=\frac{E}{E_{s}\overline{\rho}} \\
%\end{equation}

%\begin{equation} 
%\label{eq:G_norm}
%G_{norm}=\frac{G}{G_{s}\overline{\rho}} \\
%\end{equation}

%\begin{equation} 
%\label{eq:K_norm}
%K_{norm}=\frac{K}{K_{s}\overline{\rho}} \\
%\end{equation}

\subsubsection{Anisotropic Designs}
The vast majority of architectures in the dataset exhibit elastic anisotropy (Figure \ref{fig:property_overview}b). Notably, a subset of these anisotropic structures attains stiffness levels comparable to well-established lattice benchmarks such as the cubic foam. For instance, the anisotropic architecture shown in Figure \ref{fig:extreme}b achieves $E_{norm}=0.80$ at a relative density of $\overline{\rho}=0.49$, closely matching the performance of the cubic foam reported by Berger et al. \cite{berger2017mechanical}, which exhibits $E_{norm}=0.78$ at the same relative density. 

Within this broader anisotropic population, a distinct subset of highly anisotropic architectures emerges, characterized by either strong shear stiffness in the {100} planes ($Z \ge 20$), or dominant shear stiffness in the {110} planes ($Z \le 0.05$). For these extreme cases, a counterintuitive trend is observed: as relative density decreases, $E_{norm}$ increases (Figure \ref{fig:extreme}a), indicating increasingly efficient material utilization. Although these highly anisotropic structures do not reach the peak stiffness levels observed in moderately anisotropic geometries, they exhibit additional functional advantages. In particular, many display near-zero Poisson’s ratio, with a small fraction entering the auxetic regime (Figure \ref{fig:extreme}b), making them attractive for applications such as tissue engineering \cite{soman2012three}.

\subsubsection{Isotropic–Auxetic Designs}
Anisotropy can be advantageous when the loading state is well defined and predominantly aligned with a single direction. However, many load-bearing applications involve uncertain, multi-axial, or rapidly varying loading conditions, where an isotropic mechanical response is essential to ensure consistent performance independent of orientation. Although auxetic metamaterials have been widely investigated for their ability to provide enhanced energy absorption and improved shear resistance, characteristics that make them particularly attractive for biomedical applications \cite{sun2024auxetic}, achieving isotropic-auxetic structures remains a significant challenge. Isotropic-auxetic structures are rarely reported in the literature \cite{li2023three}, and often require topology optimization techniques to achieve \cite{meier2024obtaining}. In contrast, our dataset yields 442 isotropic-auxetic architectures that exhibit Poisson’s ratios as low as $\nu = -0.76$ with $0.05 \le \overline{\rho} < 0.40$. This performance is comparable to the topologies reported by Chen et al. \cite{chen2018computational}, although it remains below the extreme values near $\nu \approx-1$ demonstrated by Li et al. \cite{li2023three}.

\subsubsection{Ultra-stiff Isotropic Designs near Theoretical Bounds}
\label{subsec:ultra_stiff_isotropic}
In addition to the auxetic families, our dataset contains 66 isotropic-optimal architectures exhibiting exceptionally high effective stiffness with $0.44 \le \rho \le 0.50$. As shown in Figure \ref{fig:extreme}a,c,d, these structures exhibit among the highest elastic moduli, reaching up to 93\% of the Hashin–Shtrikman upper bound in terms of $\overline{E}_{\mathrm{mean}}/E_{\mathrm{HSU}}$. The structure obtaining an average directional Young's modulus up to 93\% of the Hashin-Shtrikman upper bound is visualized and pointed to the orange diamond in Figure \ref{fig:extreme}a. This performance is comparable to values reported in the literature, such as Zhao et al. \cite{zhao2025near}, who achieved over 98\% of the Hashin–Shtrikman bounds in the least favorable loading direction for structures with relative densities ranging from 0.2–1. Such highly stiff lightweight structures can serve as promising candidates for load-bearing applications in aerospace \cite{du2020laser}.

Furthermore, while the symmetry guided design framework generates a large number of previously unexplored architectures, it also recovers well known lattice topologies. In particular, the isotropic-optimal structures shown in Figure \ref{fig:extreme}a,c pointed to the orange diamond markers, exhibit strong geometric similarities to the Schwarz diamond lattice and the octet foam, respectively.

The Schwarz diamond-like structure identified here belongs to space group $Fm\bar{3}c$, with a relative density of $\overline{\rho}=0.49$ and a normalized Young’s modulus of $E_{norm}=0.65$. This value exceeds the normalized stiffness reported for the Schwarz diamond lattice studied by Alagha et al. \cite{alagha2024mechanical}, who measured $E_{norm} \approx 0.54$ at a similar relative density of $\overline{\rho}=0.5$. Similarly, the octet foam-like structure recovered by our framework belongs to space group $Fm\bar{3}m$ and achieves $E_{norm}=0.58$ at $\overline{\rho}=0.49$, which is comparable to the octet foam analyzed by Berger et al. \cite{berger2017mechanical}, who reported $E_{norm}=0.53$ at $\overline{\rho}=0.53$. Notably, the octet foam in the study by Berger et al. exhibits anisotropic behavior, whereas the structure identified in our dataset approaches isotropic performance.

\subsubsection{Pentamode “metafluid” Designs}
\begin{figure}[htbp]
\centering
\includegraphics[width=0.95\textwidth]{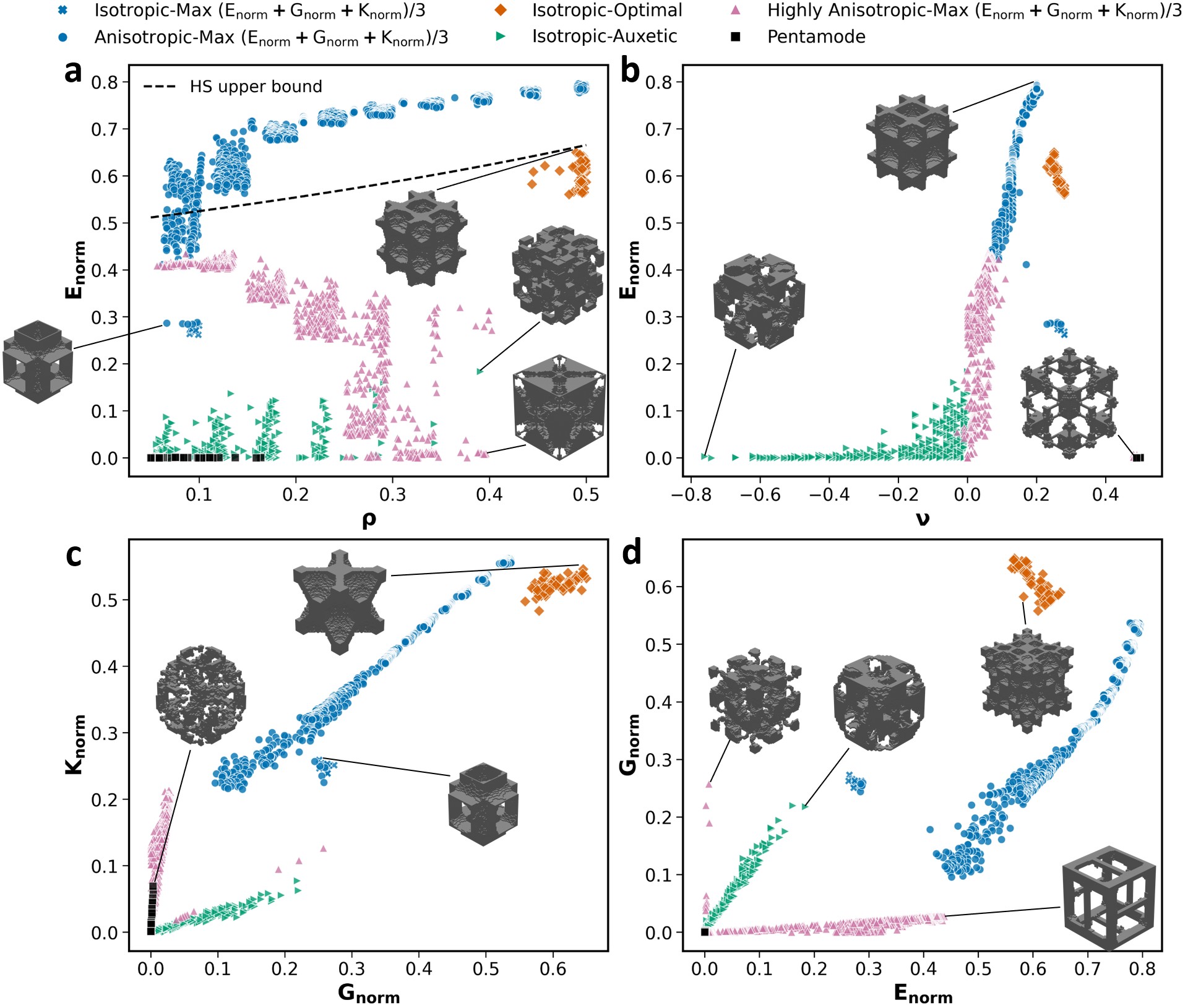}
\caption{The elastic properties of various classes of extreme structures. (a) Relative density plotted against normalized Young's modulus. (b) Poisson's ratio plotted against normalized Young's modulus. (c) Normalized Young's modulus plotted against normalized bulk modulus. (d) Normalized Young's modulus plotted against normalized shear modulus.}
\label{fig:extreme}
\end{figure}

We also report pentamode metamaterials in our dataset, which are structures engineered to exhibit extremely high bulk modulus but negligible shear resistance so that it is rigid under volumetric loads while offering almost no resistance to shape change, as such they are also referred to as 'meta-fluids' \cite{wu2021topological}. When K is extremely large compared to the shear modulus, the Poisson's ratio $\nu = \frac{3 - 2(G/K)}{2(G/K) + 6}\approx0.5$. Pentamode metamaterials can be used for elastomechanical cloaking, which manipulates the elastic response around a protected region, effectively rendering the enclosed object mechanically "unfeelable" \cite{buckmann2014elasto}\cite{fielding2024simple}. Figure \ref{fig:extreme}c, d illustrate this characteristic "fluid-like" behaviour, with the pentamode structures exhibiting a relatively high bulk modulus compared to the almost vanishing shear resistance. Within this dataset we produced 72 pentamode structures with $0.05 \le \rho \le 0.16$. An exceptional pentamode structure with a high K/G ratio is visualized in Figure \ref{fig:extreme}b pointed to the black square. The eigenvalues of the stiffness matrix for this structure is $[0.09, 0.09, 0.09, 0.22, 0.22, 27.28]$, where the last value is at least two orders of magnitude greater than the others. The eigenvector $e\approx [0.58, 0.58, 0.58, 0, 0, 0]$ corresponding with the highest eigenvalue shows three equal normal strains and three zero shear strains, verifying that the dominant deformation mode is hydrostatic. The shear modulus $G=0.08$ and bulk modulus $K=13.30$ of this visualized structure produce a ratio of $K/G\approx166$. This ratio is comparable to other values reported in the literature. For example, Lumpe and Stankovic \cite{lumpe2021exploring} identified a base unit cell with $K/G\approx113$, and Wu et al. \cite{wu2021topological} achieved ratios up to $K/G\approx212$. Although, substantially higher values have also been demonstrated, such as Schittny et al. \cite{schittny2013elastic} who reported ratios near $K/G\approx1000$, and Lumpe and Stankovic \cite{lumpe2021exploring} who further increased their base unit-cell performance to $K/G\approx670$ by replacing its cylindrical members with biconical members.

Although our structures do not always reach the most extreme values reported in prior work, many of those studies either employ targeted topology optimization or systematically tailor specific geometric motifs to amplify particular mechanical responses. In contrast, our dataset is generated solely through cubic crystal symmetry operations, without any task specific tuning. While existing studies report specific classes of high performance architectures, it is comparatively rare for a single generative procedure to produce such a diverse set of extreme material families within one large-scale dataset. These findings indicate that symmetry constrained generation is capable of yielding rare, high-performance geometries, underscoring the richness of cubic symmetry design spaces. We effectively show that symmetry-guided design can unlock a treasure trove of high-performance structures across multiple property axes simultaneously.

\FloatBarrier

\FloatBarrier

\subsection{Influence of Symmetry on Stiffness Properties}

To evaluate the significance of relationships between mechanical properties and cubic crystal symmetry descriptors, namely space group, lattice type, and point group, we employed the nonparametric Kruskal-Wallis test ($\alpha =0.05$). The lattice types correspond to the three cubic Bravais configurations: simple cubic (SC), face-centered cubic (FCC), and body-centered cubic (BCC). While the point groups belong to the cubic crystal classes: tetartoidal, diploidal, gyroidal, hextetrahedral, and hexoctahedral systems. The Kruskal-Wallis test was used to assess whether the rank sums and distributions of continuous mechanical properties ($E_{norm}$, $G_{norm}$, $K_{norm}$, $\nu$, and $Z$) differ significantly across symmetry descriptors. This nonparametric approach was selected given the right-skewed distributions observed across the mechanical property space.
For all property to symmetry variable comparisons, the Kruskal-Wallis tests yielded p-values that approached zero. This is expected given the extremely large sample sizes, where even small differences in the distributions become statistically significant. Therefore, statistical significance alone is insufficient to assess the strength of these relationships. To better capture their practical importance, we computed corresponding effect sizes ($\epsilon^2$), which quantify the magnitude of the association between groups, alongside visual assessment of the underlying distributions. The interpretation of the $\epsilon^2$ values followed guidelines suggested by Mangiafico \cite{mangiafico2016summary}.

\begin{figure}[htbp]
\centering
\includegraphics[width=0.9\textwidth]{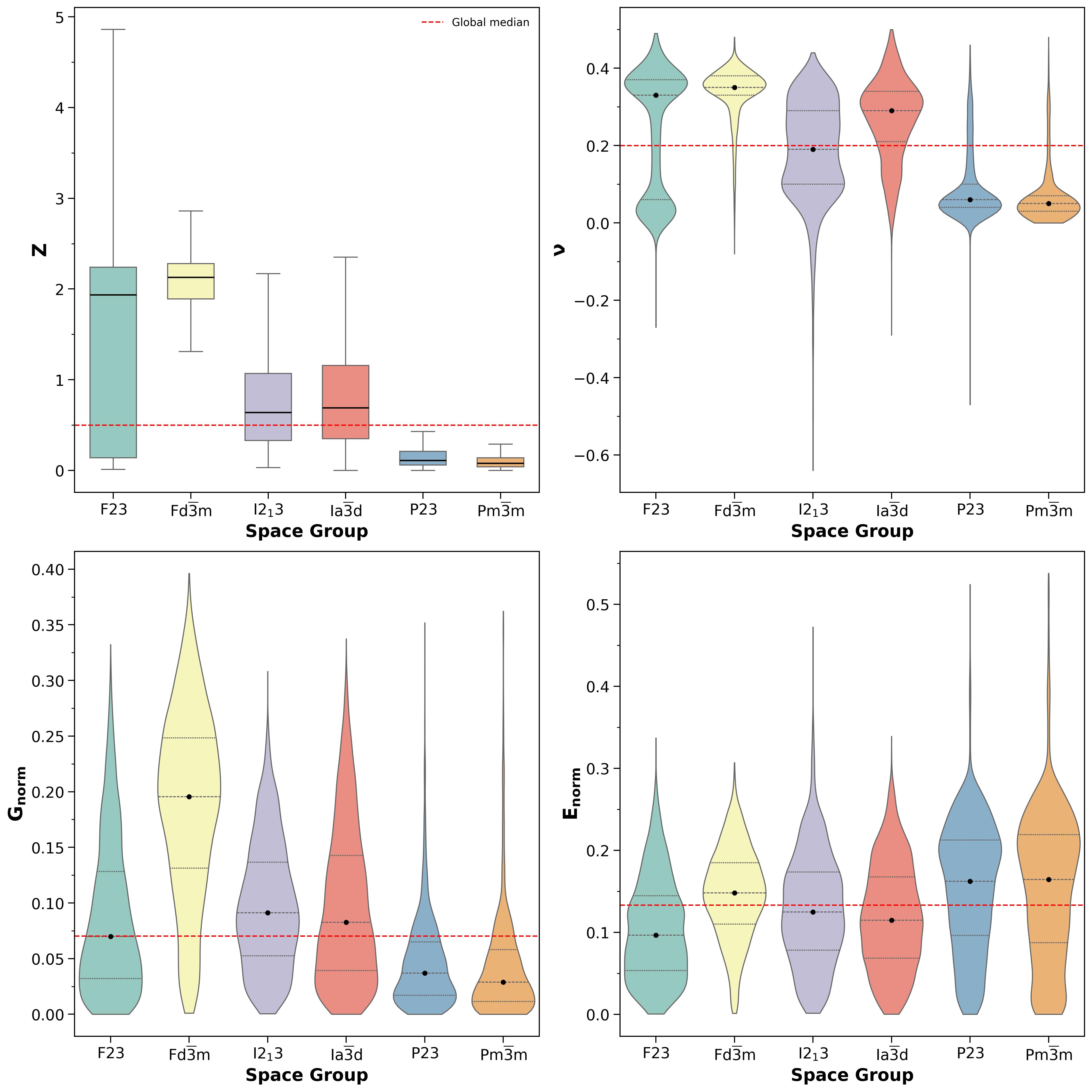}
\caption{Distributions of elastic moduli and Zener anisotropy across representative cubic space groups within the low relative density regime, balanced by random down sampling ($\rho \in [0.05,0.2]$). For clarity, only a subset of representative cubic space groups are shown. The selected space groups span multiple Bravais lattices and symmetry classes and exhibit distinct property distributions. Space groups 196, 227, 199, 230, 195, and 221 are labeled using Hermann–Mauguin notation. The red dashed lines indicate the global median across all cubic space groups visualized. (a) Zener anisotropy ratio $Z$, (b) Poisson’s ratio $\nu$, (c) normalized shear modulus $G_{norm}$, and (d) normalized Young’s modulus in the $\langle 100 \rangle$ directions $E_{norm}$ are shown for select space groups. 
}
\label{fig:distributions}
\end{figure}

From Table \ref{table:low_dens_hypothesis}, the Kruskal-Wallis tests indicate that space group exhibits a large effect on both Zener anisotropy and Poisson’s ratio, as reflected by the corresponding effect sizes. Although Bavais lattice type and point group symmetry, when considered independently, exert a relatively small influence on mechanical properties, varying combinations of them leads to pronounced variations in both the central tendency and spread of the property distributions. This is evident in Figure \ref{fig:distributions}a,b, where distinct shifts in medians and violin distribution shapes emerge across different space groups. Notably, higher symmetry does not always imply a differing mechanical response. For example, the space groups $P23$ and $Pm\overline{3}m$ display similar property distributions despite differing point group symmetries. This similarity can possibly be attributed to the subgroup relationship of $P23$ within $Pm\overline{3}m$. Although, the space groups $F23$ and $Fd\overline{3}m$ exhibit more pronounced distributional differences across the same properties, despite $F23$ being a subgroup of $Fd\overline{3}m$. 

Among the mechanical properties tested against Bravais lattice type, Zener anisotropy stands out as it seems to have a comparatively larger effect size than the other tests Bravais lattice tests conducted, although it remains at the upper bound of what is typically classified as a small effect. Nevertheless, inspection of the underlying distributions reveal that FCC lattices generate structures spanning a broader range of Zener anisotropy ratios, largely between $0-2$, while SC lattices predominately give rise to more anisotropic responses with stronger normal stiffness $Z < 0.5$ (Supplementary Information Section 2).

Furthermore, space groups $Ia\overline{3}$, $F432$, $F\overline{4}3c$, and $Fm\overline{3}c$ each exhibit auxetic behavior in at least 10\% of their structures within the low relative density regime. The latter three are closely related through subgroup-supergroup relationships, suggesting possibly that specific symmetric features within this symmetry family promote geometric configurations conducive to auxetic responses. Interestingly, despite not belonging to this subgroup hierarchy, $Ia\overline{3}$ yields the highest proportion of auxetic structures, with approximately 12\% of its geometries exhibiting negative Poisson’s ratio, demonstrating that multiple symmetry mechanisms may give rise to auxetic behavior. Additionally, we found that of the 66 isotropic-optimal structures present in the dataset, 53 are FCC and 32 belong to the $m\overline{3}m$ point group, possibly suggesting that these symmetry parameters are particularly favorable for realizing architectures with high stiffness.

Collectively, these results demonstrate that mechanical properties such as the Zener anisotropy ratio and Poisson’s ratio are strongly governed by structural symmetry. In contrast, the magnitude of the shear modulus exhibits a weaker dependence on crystallographic structure, with Young’s modulus and bulk modulus showing even less sensitivity. This distinction highlights that the space group primarily modulates the directionality of mechanical behavior rather than its overall stiffness scale, showcasing symmetry as a key determinant of mechanical diversity in low relative density architected materials. For additional details on the symmetry to property distributions see supplementary information Section 2.

\FloatBarrier

\begin{table}[htbp]
\centering
\caption{Summary of hypothesis test results for mechanical properties and symmetry descriptors across low relative density structures ($\rho \in [0.05, 0.2]$). For the Kruskal-Wallis tests, the null hypothesis was that the distributions of the corresponding mechanical property were identical across space group, Bravais lattice and point group.}
\label{table:low_dens_hypothesis}
\scriptsize
% \begin{tabular}{llrrrrrl}
\begin{tabularx}{\textwidth}{lXrrrrrl}
\toprule
\textbf{Variables Compared} & \textbf{Statistic (H)} & \textbf{Degree of Freedom} & \textbf{N} & \textbf{$p$-value} & \textbf{Effect Size ($\epsilon^2$)} & \textbf{Interpretation} \\
\midrule
Space Group $\times$ $E_{norm}$ & 40018.2 & 35 & 500{,}663 & $p < .001$ & 0.08 & Moderate \\
Space Group $\times$ $G_{norm}$ & 93740.4 & 35 & 500{,}663 & $p < .001$ & 0.19 & Moderate \\
Space Group $\times$ $K_{norm}$ & 41482.5 & 35 & 500{,}663 & $p < .001$ & 0.08 & Moderate \\
Space Group $\times$ Z & 169858.8 & 35 & 500{,}663 & $p < .001$ & 0.34 & Large \\
Space Group $\times$ $\nu$ & 163300.8 & 35 & 500{,}663 & $p < .001$ & 0.33 & Large \\
Bravais Lattice $\times$ $E_{norm}$ & 9264.9 & 2 & 500{,}663 & $p < .001$ & 0.02 & Small \\
Bravais Lattice $\times$ $G_{norm}$ & 18613.3 & 2 & 500{,}663 & $p < .001$ & 0.04 & Small \\
Bravais Lattice $\times$ $K_{norm}$ & 8770.3 & 2 & 500{,}663 & $p < .001$ & 0.02 & Small \\
Bravais Lattice $\times$ Z & 34643.1 & 2 & 500{,}663 & $p < .001$ & 0.07 & Small \\
Bravais Lattice $\times$ $\nu$ & 19174.8 & 2 & 500{,}663 & $p < .001$ & 0.04 & Small \\
Point Group $\times$ $E_{norm}$ & 5028.9 & 4 & 500{,}663 & $p < .001$ & 0.01 & Small \\
Point Group $\times$ $G_{norm}$ & 8796.3 & 4 & 500{,}663 & $p < .001$ & 0.02 & Small \\
Point Group $\times$ $K_{norm}$ & 7322.4 & 4 & 500{,}663 & $p < .001$ & 0.02 & Small \\
Point Group $\times$ Z & 14230.0 & 4 & 500{,}663 & $p < .001$ & 0.03 & Small \\
Point Group $\times$ $\nu$ & 10224.4 & 4 & 500{,}663 & $p < .001$ & 0.02 & Small \\
\bottomrule
\end{tabularx}
\end{table}

\FloatBarrier

\FloatBarrier
\subsection{Finite-element results}\label{subsec:fea}
\begin{figure}[htbp]
\centering
\includegraphics[width=0.95\textwidth]{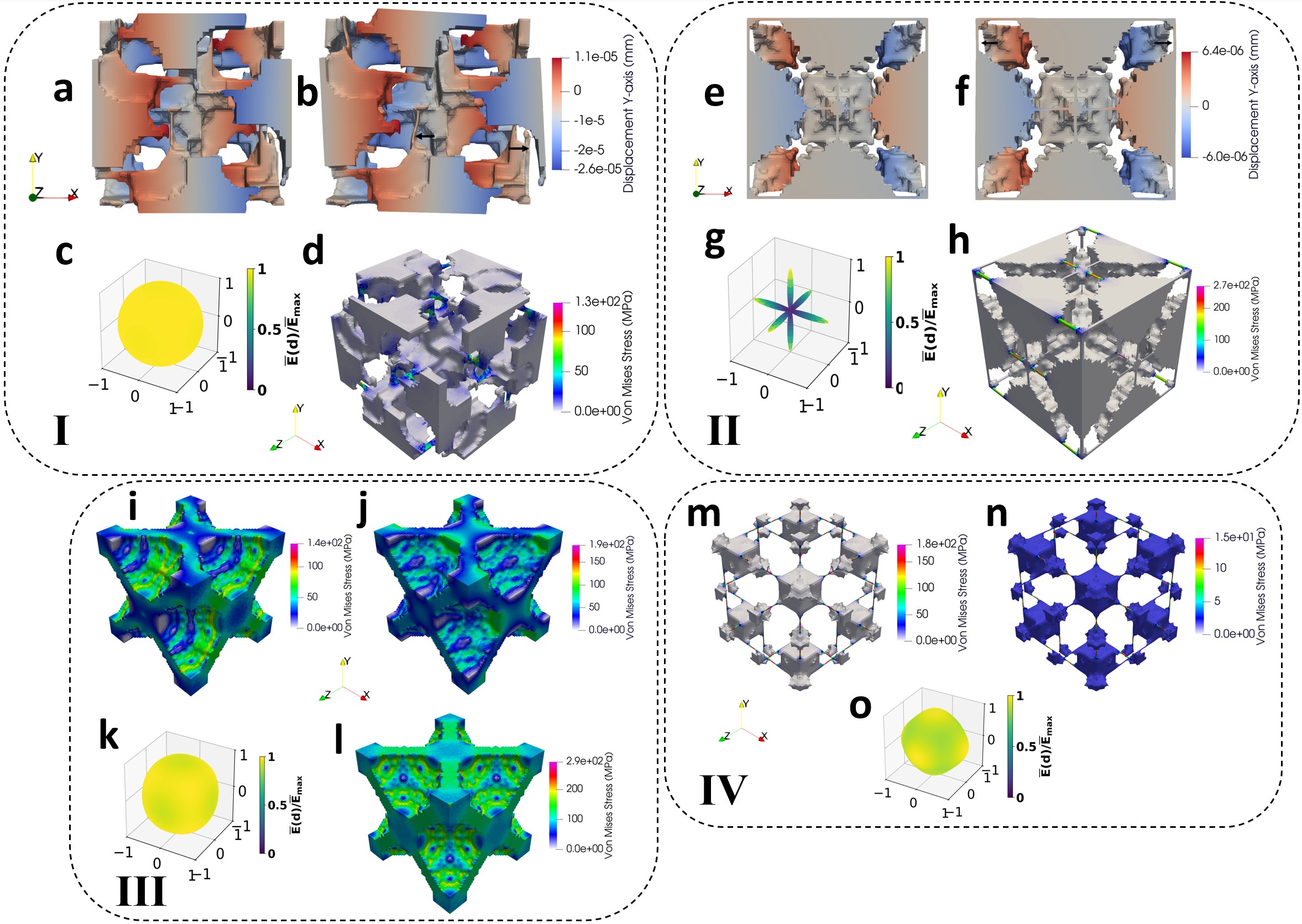}
\caption{FEA analysis of an isotropic-auxetic sample, highly anisotropic, isotropic-optimal, and pentamode sample. The stiffness tensor plot is presented to illustrate the degree of directional dependence or isotropy in the elastic response of these structures, where $\overline{E}(d)$ is the directional effective Young's modulus and $\overline{E}_{max}$ is the maximum directional Young's modulus value. (a) Left-hand side view of an isotropic-auxetic sample colored by transverse displacement in the y-axis field under uniaxial tensile loading along the x-direction (b) Warped left-hand side view of an isotropic-auxetic sample, demonstrating lateral expansion under uniaxial tensile loading along the x-direction. (c) Isotropic-auxetic stiffness tensor plot (d) Isotropic-auxetic Von Mises stress field under uniaxial tensile loading. (e) Left-hand side view of a highly anisotropic sample colored by transverse displacement in the y-axis field under uniaxial tensile loading along the x-direction (f) Warped left-hand side view of a highly anisotropic sample, demonstrating no lateral expansion under uniaxial tensile loading along the x-direction. (g) Highly anisotropic stiffness tensor plot (h) Highly anisotropic Von Mises stress field under uniaxial tensile loading. (i) Isotropic-optimal Von Mises stress field under uniaxial tensile loading. (j) Isotropic-optimal Von Mises stress field under shear loading. (k) Isotropic-optimal stiffness tensor plot (l) Isotropic-optimal Von Mises stress field under hydrostatic loading. (m) Pentamode Von Mises stress field under hydrostatic loading. (n) Pentamode Von Mises stress field under shear loading. (o) Pentamode stiffness tensor plot}
\label{fig:fea}
\end{figure}

Figure \ref{fig:fea}a,b shows the left-hand side view of the isotropic-auxetic structure with Poisson's ratio $\nu = -0.76$. Under uniaxial tensile loading along the x-axis, the structure undergoes concurrent expansion in the transverse y-axis. The corresponding Von Mises stress fields shown in Figure \ref{fig:fea}d indicate that the deformation response is governed predominantly by the network of internal ligaments. Elevated stress concentrations are localized along these ligament segments, reflecting deformation that is characteristic of chiral-like bending mechanisms. This bending motion within the unit cell facilitates transverse expansion under axial tension, giving rise to the observed auxetic response.

The highly anisotropic architecture shown in Figure \ref{fig:fea}e,f,h exhibits a markedly greater stiffness under uniaxial loading than under shear and displays a near-zero Poisson’s ratio ($\nu = 0.01$). As a result, the structure undergoes negligible transverse deformation when loaded uniaxially along the x-axis (Figure \ref{fig:fea}f). Examination of the Von Mises stress fields under uniaxial tensile loading further reveals that load is predominantly carried by the slender truss members aligned with the loading direction, along which stresses are concentrated. In contrast, members oriented perpendicular to the loading axis remain largely unstressed, indicating minimal load transfer in the transverse directions. This highly directional load-bearing mechanism suppresses lateral deformation and directly gives rise to the observed near-zero Poisson’s ratio.

In section \ref{subsec:ultra_stiff_isotropic} we compare the the octet-foam to the isotropic-optimal structure visualized in Figure \ref{fig:extreme}c. Figure \ref{fig:fea}e-g show the von Mises stress fields of the isotropic-optimal topology under uniaxial, shear, and hydrostatic loading. Across all three loading conditions, the stress distributions remain roughly uniform, with no pronounced stress localization. Whereas with the octet foam, while it generally has uniform stress, it exhibits localized stress concentrations at the junctions of its plate members \cite{liu2021mechanical}, with those concentrations arising from the abrupt change in load path. This brings out the key distinction between our isotropic-optimal structure and the traditional octet-foam. Rather than being composed of plates intersecting with sharp junctions, the isotropic-optimal topology forms smooth, continuous load-bearing pathways. This geometric continuity facilitates gradual stress redistribution, mitigates stress amplification at junctions, and promotes more uniform stress fields under multi-axial loading. As a result, the geometric efficiency is enhanced relative to plate-based octet-foam architectures.

Lastly, examination of the stress contours for the pentamode structure under hydrostatic loading (Figure \ref{fig:fea}m) and shear loading (Figure \ref{fig:fea}n) reveals that load is primarily concentrated in the slender truss members. Under hydrostatic loading, these members experience predominantly axial deformation, exhibiting a stretch-dominated response that efficiently transmits volumetric stress. A comparison of the corresponding stress scales further indicates that the structure sustains substantially higher stress levels under hydrostatic loading than under shear loading. This pronounced contrast is consistent with the defining characteristic of pentamode metamaterials, where this particular structure achieves $K/G\approx166$.

\FloatBarrier

\FloatBarrier

\subsection{3D-CNN Surrogate Model}\label{subsec:ml-framework}

\begin{figure}[htbp]
\centering
\includegraphics[width=0.9\textwidth]{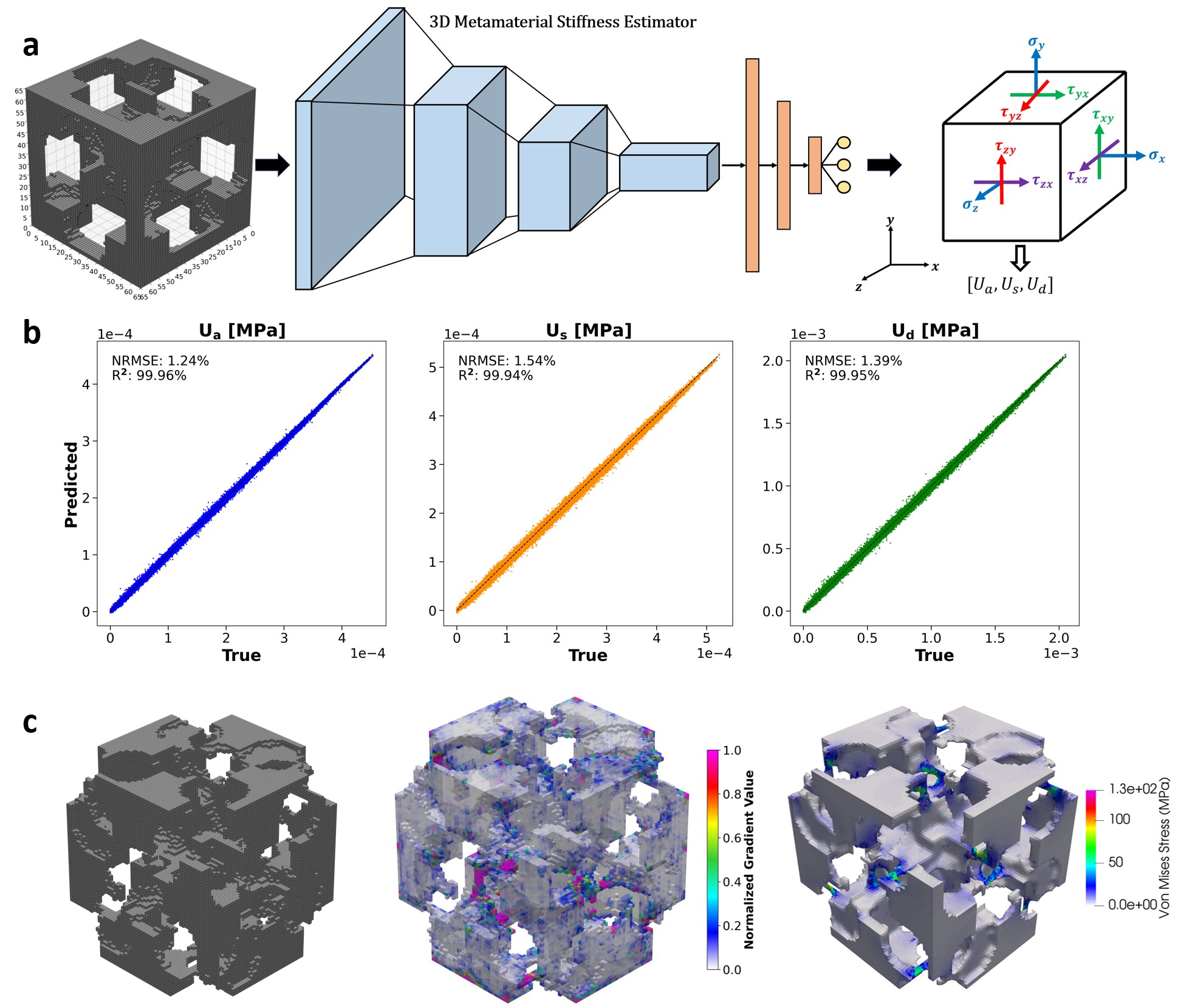}
\caption{Overview and performance of the CNN pipeline used for stiffness prediction. (a) CNN pipeline. (b) Accuracy of predictions on uniaxial, shear and hydrostatic strain energy densities. (c) Explainability analysis on an isotropic-auxetic structure, with the gradients of the saliency map normalized from 0 to 1.}
\label{fig:cnn}
\end{figure}

CNNs were originally developed for computer vision and have demonstrated remarkable success in various image-related tasks due to their ability to effectively capture spatial features and patterns \cite{krizhevsky2017imagenet}\cite{he2016deep}\cite{wang2016cnn}\cite{gu2018recent}. In recent years, researchers have increasingly leveraged CNNs for predicting mechanical properties of materials, capitalizing on their ability to capture complex, non-linear structure-property relationships within datasets \cite{Wilt2020}\cite{rao2020three}\cite{zhao2025machine}. Once trained, these models can rapidly infer material behavior from geometry, offering a computationally efficient surrogate to traditional FEM simulations while maintaining high predictive accuracy. Here we leverage the full database of approximately 1.95 million vectorized topologies to train and evaluate a convolutional neural network containing $\sim$7.7 million parameters. The targets are the strain energy densities under uniaxial, shear and hydrostatic loading ($U_a$, $U_s$, $U_d$). Built using Tensorflow and Keras, the model architecture developed in this work is inspired by ResNet architecture, which employs residual connections to improve gradient flow during backpropagation and mitigate vanishing gradient issues in deep networks \cite{he2016deep}. The model input tensor has the shape (N,65,65,65,1) where N denotes the batch size, 65×65×65 corresponds to the voxel grid dimensions along the x, y, and z axes, and the final dimension represents the single input channel. The network starts with a convolutional layer comprising 32 filters and kernel size of 3, followed by a batch normalization layer. Batch normalization layers are added to improve training stability \cite{ioffe2015batch}. This is followed by four residual blocks, where the number of filters increases progressively from 32 to 256, with the starting convolutional layer of each residual block having a stride of 2 to perform spatial down sampling. After the convolutional and residual operations, the feature maps are flattened and passed through fully connected layers, ultimately producing three linear outputs corresponding to the predicted uniaxial, shear, and hydrostatic strain energy densities (Figure \ref{fig:cnn}a). The model was trained for 10 epochs until convergence with an 80/10/10 randomized train-validation-test split using the Adam optimizer, a learning rate of 1e-3, and a batch size of 32.
The CNN achieves at least $R^2$=99.9\% and a NRMSE less than 2\% (normalized by the interquartile range of the true values) across all targets when evaluated on the hold-out test set. These results demonstrate the CNN’s ability to accurately approximate elastic properties of complex metamaterial architectures, enabling rapid high-throughput screening prior to FEM validation. For further details regarding the full model architecture and training scheme refer to supplementary information Section 3.

Furthermore, we perform a post-hoc saliency analysis of the CNN predictions for the isotropic-auxetic structure discussed in section \ref{subsec:fea}. By comparing regions of high gradient magnitude in the CNN’s prediction of the Young’s modulus along the $\langle 100 \rangle$ directions with the von Mises stress fields obtained from uniaxial loading, a clear qualitative correspondence between the two can be observed (Figure \ref{fig:cnn}c). Specifically, ligament regions identified with high gradient magnitudes by the network tend to spatially coincide with ligaments experiencing elevated stress, suggesting that the CNN has learned to associate mechanically critical geometric features with its predictions. Regions exhibiting elevated gradient magnitudes indicate areas to which the model is particularly sensitive when generating predictions for this structure. Despite this qualitative alignment, saliency maps should not be interpreted as ground-truth indicators of mechanical relevance. Gradient-based attribution methods should be used as heuristic tools as they may not always correspond directly to physically meaningful stress or strain distributions. In some cases, salient regions may be distributed in a manner that is challenging to interpret and may not directly correspond to regions of mechanical significance. Nevertheless, when used judiciously, saliency analysis provides a valuable complementary tool for probing the internal representations learned by the network and for generating ideas about influential geometric motifs. These results highlight the potential of deep learning not only as a surrogate predictor of mechanical properties, but also as a means of uncovering structure-property relationships that merit further targeted investigation. For more details on the implementation of these saliency maps, see supplementary information Section 4.

\FloatBarrier

\section{Conclusion}
Here we have proposed a cubic crystal symmetry based design approach of metamaterial unit cells that go beyond the commonly studied strut, plate, and shell-based lattices, creating a novel dataset of nearly two million unit cells. Our primary focus has been investigating how cubic space-group symmetries govern elastic mechanical response, showing that specific symmetry operations strongly influence anisotropy and Poisson's ratio behaviour. Using our symmetry-based generation algorithm, we demonstrate the ability to produce various families of extreme metamaterial topologies, including isotropic-optimal, isotropic-auxetic, highly anisotropic and pentamode structures. The comprehensive catalog of metamaterial families presented in this dataset provides researchers and engineers with access to a diverse design space spanning isotropic and anisotropic responses, as well as both stiff and compliant architectures. This breadth enables systematic selection and tailoring of unit-cell geometries to meet application-specific performance requirements, including biomedical devices, aerospace structures, and elastomechanical cloaking. Beyond linear elasticity, the geometries contained in this dataset offer a foundation for future investigations into multifunctional behavior, such as coupled electrical and thermal transport, as well as strength, toughness, and other nonlinear mechanical responses.
\setlength{\parskip}{0.5em}

Complementing the dataset, the trained deep learning surrogate model enables rapid and scalable screening of metamaterial topologies. This framework can extend beyond forward property prediction to support inverse design. By navigating the learned structure-property landscape, the model can be used to generate previously unseen unit-cell architectures with stiffness values that surpass those observed in the original dataset, highlighting its potential as a practical tool for data-driven metamaterial discovery and optimization.

\FloatBarrier

\section{Methods}

\subsection{Topology Generation}
A high throughput generative framework was developed to synthesize architected materials constrained by the 36 cubic space groups, corresponding to space groups 195 to 230. Inspired by crystallographic symmetry principles \cite{Mao2020}, unit cells were discretized into binary voxel grids of resolution $65^3$, initialized from a fully connected solid configuration. Controlled stochastic erosion was then applied to achieve target relative densities within the range $\overline{\rho} \in (0, 0.5]$, while strictly enforcing space group specific symmetry operations and periodic boundary conditions. Over two million symmetry-preserving lattices were initially generated, then structures with $\overline{\rho} < 0.05$ were filtered out due to potential meshing and numerical stability issues. In addition, any unit cells that violated periodic boundary conditions were removed, yielding a final dataset of approximately $1.95$ million valid structures. Through this generative framework, a broad spectrum of topologies is produced, spanning bending-dominated and stretching-dominated architectures, as well as both open-cell and closed-cell configurations. While some geometries bear resemblance to well-established lattice designs, many are entirely novel, featuring complex sheet-like load-bearing walls or encapsulated void morphologies. The details of the topology generation algorithm are outlined in the supplementary information Section 1.

\subsection{Meshing}
The methodology for generating a mesh from a 3D binary array representing a metamaterial geometric structure involves several key steps. Firstly, the MATLAB iso2mesh toolbox was utilized to extract the bounding mesh surface of the metamaterial \cite{fang2009tetrahedral}. Subsequently, the meshing algorithm, implemented through the cgalmesh binary executable \cite{jamin2015cgalmesh}, employed an oracle to handle surface intersections and compute intersection points using restricted Delaunay triangulation principles \cite{boissonnat2005provably,rineau2007generic}, ensuring a periodic and homeomorphic surface mesh. Logical masks were then applied to identify surface elements within the unit cell's bounding box, followed by Laplacian smoothing to enhance mesh robustness. Using the refined surface mesh, tetrahedral elements were generated to represent the metamaterial's 3D volume using TetGen \cite{hang2015tetgen}, resulting in a quadratic tetrahedral mesh. Finally, the processed mesh data, including nodes and elements were saved in ABAQUS input file format (.inp) and later converted to a Code-Aster mesh file (.med) using meshio \cite{meshio}. 

\subsection{Finite-Element Modeling}
Finite element simulations were conducted using Code Aster (v14.06) on CPU resources provided by the Digital Research Alliance of Canada. The constituent material of each unit cell was modeled as isotropic, linear elastic steel with Young’s modulus $E_s = 205{,}000$ MPa and Poisson’s ratio $\nu_s = 0.29$. Unit cell models employed standard periodic boundary conditions \cite{danielsson2002three}. Uniaxial, shear, and hydrostatic loading cases were applied to compute the corresponding strain energy densities ($U_a$, $U_s$, $U_d$). These quantities were used to extract the three independent stiffness tensor components ($C_{11}$, $C_{12}$, $C_{44}$), from which the effective elastic constants ($E$, $G$, $K$, $\nu$) and the Zener anisotropy ratio $Z$ were subsequently derived.

\subsection{Explainability}
To gain insight into the CNN’s predictive behavior, we employ 3D saliency maps to identify which voxels within each unit cell most strongly influence the model’s predictions, thereby providing a measure of input sensitivity. Saliency maps are computed using the SmoothGrad method, a gradient-based explainability technique that enhances saliency maps by reducing gradient noise through input perturbation \cite{smilkov2017smoothgrad}. Specifically, SmoothGrad generates multiple noisy copies of the input by adding Gaussian noise, computes the gradient of the model output with respect to each copy, and averages the resulting gradients to highlight regions that consistently contribute to the prediction. Following the work of Smilkov et al. \cite{smilkov2017smoothgrad} we use average between 50 noisy copies and set the variance of the Gaussian noise to 0.2. This procedure yields a 65×65×65 voxel-level saliency map for each unit cell. We then apply a mask to the saliency map to retain only gradients corresponding to voxels where the structure exists, allowing us to isolate the attributions that are physically meaningful. Additionally, to prevent a small number of extreme gradient values from dominating the visualization, saliency values are clipped at the 99th percentile.

\subsection{Experimentation}

\begin{figure}[htbp]
\centering
\includegraphics[width=0.95\textwidth]{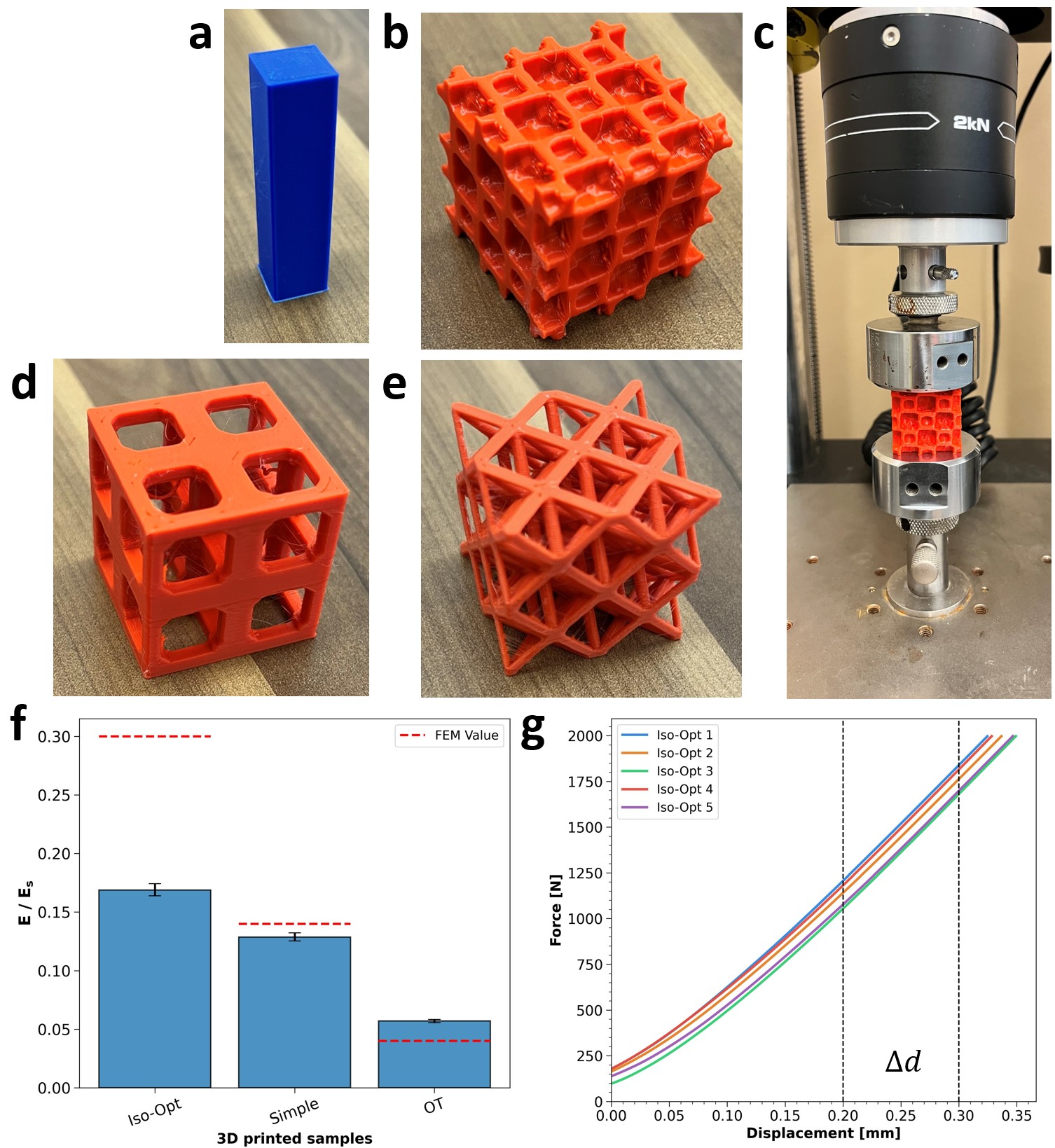}
\caption{Experimental fabrication and mechanical testing of FFF printed architected materials. (a) The bulk sample; (b) Isotropic-optimal sample printed as a 2x2x2 array with 1.5cm unit cells. (c) Quasi-static uniaxial compression testing apparatus, loading the isotropic-optimal sample. d) Simple sample printed as a 2x2x2 array with 1.5cm unit cells. e) Octet-truss sample printed as a 2x2x2 array with 1.5cm unit cells. f) Comparison of the average relative stiffness values of all the lattice samples tested. Error bars represent the worst case bounds based on the extrema of the measured stiffness and compression modulus. g) Force-displacement curves of the five isotropic-optimal samples, $\Delta d$ denotes the displacement range along the linear region used to compute the Young's modulus.}
\label{fig:experimental}
\end{figure}

To validate our finite-element results, we fabricated three representative metamaterial architectures and performed uniaxial compression tests to extract their effective Young's modulus. Two structures from the dataset were manufactured using fused filament fabrication (FFF) on a Prusa Mk4 with PLA+: a simple geometric lattice (Figure \ref{fig:experimental}d) and an isotropic-optimal design (Figure \ref{fig:experimental}b), each with the same 30 × 30 × 30 mm dimensions. To establish bulk properties for the parent materials, we fabricated a dedicated set of 10 bulk specimens (Figure \ref{fig:experimental}a). The bulk specimens were produced with the standardized dimensions of 12.7 x 12.7 x 50.8 mm as outlined in the ASTM D695 standards \cite{ASTMStandardD695232023}. All of the samples printed using PLA+ were oriented such that their layer lines would be perpendicular to the loading axis. As an additional validation step, leveraging the octet-truss (OT), one of the most extensively studied metamaterial architectures with well-established theoretical and experimental stiffness responses, we fabricated five OT lattices (Figure \ref{fig:experimental}e). These samples were then characterized under uniaxial compression using an Instron 5848 microtester, as shown in Figure \ref{fig:experimental}c. A constant displacement rate of 1.3 mm/minute was applied between two parallel plates with the load measured by a 2,000N capacity load cell. Force and displacement data were continuously recorded, and then shifted along the x-axis to exclude most of the displacement associated with the initial settling phase, producing the final force-displacement curves. To determine the Young's modulus of each sample, the applied load was converted to a nominal apparent stress by dividing by the theoretical bulk cross-sectional area (${L_{lattice}}^2$), and the strain was computed using the nominal specimen height. A consistent displacement window of 0.1 mm was applied to all samples, and the effective Young’s modulus was obtained by averaging the response computed response within this linear section. The average measured Young's modulus of all the bulk samples is 1238.5 $\pm$ 20.6 MPa, this is similar to the work of Pastor-Artigues et al. \cite{pastor2019elastic} who report an average compression modulus of 1370.2 MPa for PLA. Figure \ref{fig:experimental}f compares the relative stiffness of each lattice sample with their corresponding FEM predictions.

The isotropic-optimal lattice exhibits a relative error of 43.3\%, compared to 50.0\% for the octet-truss and 7.1\% for the simple lattice. The simple lattice shows close agreement with the FEM predictions, whereas larger discrepancies are observed for the octet-truss and the isotropic optimal architecture. Notably, despite its larger relative error, the octet-truss displays a small absolute deviation from the FEM value, with a difference of only 0.02, whereas the isotropic-optimal lattice deviates by 0.13. The larger discrepancy observed for the isotropic-optimal topology possibly suggests that this architecture may be particularly sensitive to fabrication constraints. In particular, the closed-cell, intricate internal geometry likely introduces challenges for the 3D printer, leading to geometric deviations from the idealized design and, consequently, altering the mechanical response. Furthermore, it is worth noting that we only tested for one direction, whereas a more comprehensive and thorough characterization of these lattices would require reorienting the lattices using CAD to enable loading in the [110] and [111] directions, similar to what was done in the work by Yu et al.~\cite{yu2023truss}. For more details relating to the experimental results, refer to Section 5 in the supplementary information.

\newpage
\section*{Data Availability}
The data and code will be made available on the Digital Research Alliance of Canada website.

\section*{Acknowledgements}
The authors gratefully acknowledge the financial support provided by the Natural Sciences and Engineering Research Council of Canada (NSERC) and the Digital Research Alliance of Canada.
The authors also thank Dr. Hani Naguib and Dr. Adam Pearson of the University of Toronto’s Department of Mechanical and Industrial Engineering for providing access to mechanical testing facilities. Dr. Naguib granted use of his laboratory, and Dr. Pearson provided training for the safe and proper operation of the testing equipment. In addition, the authors thank Peter Di Palma and Dr. Sanjeev Chandra of the University of Toronto’s Department of Mechanical and Industrial Engineering for 3D printing samples.

\section*{Conflict of interest}
The authors declare no conflict of interest.
\clearpage
\section*{Supporting Information}
\clearpage


\section*{Supplementary material}

# Supporting Information

**Section 1 - Topology Generation Methodology**

Figure S1 illustrates the step-by-step process of the topology generation algorithm. The process begins with an initial bulk material of size 65×65×65.

Step 1: *Random Void Seeding*: The process begins by randomly selecting and applying a number of void seed points, where the number of seeds, $N_{seed}$, is chosen within the range $1 \leq N_{seed} \leq 6$. These void seeds act as initial starting points for material removal. In this figure, a single seed point is used.

Step 2: *Symmetry Enforcement via Space Group Operations*: Next, symmetry operations are applied in accordance with the target crystallographic space group. In this example, space group 195 (P23) is enforced. These symmetry operations are applied to the initial seed points across the structure, ensuring that the design adheres to the constraints of the selected space group from the outset.

Step 3: *Sequential Erosion with Symmetry Constraints*: The void seed regions then undergo incremental erosion, where material is randomly removed along the boundaries of the seed while enforcing symmetry constraints. This sequential erosion continues until the target relative density is reached. Given the stochastic nature of the erosion, some regions of the structure may become disconnected. To prevent isolated floating objects, a connectivity enforcement step is introduced, where any detached elements are bridged back to the main structure, ensuring continuity.

Step 4: *Refinement for Periodicity and Symmetry Enforcement*: Once the erosion process concludes, the resulting structure has both periodic boundary conditions (PBCs) and full symmetry enforced.

Step 5: *Final Optimization and Defect Removal*: As a final step, periodic boundary conditions are reapplied to ensure translational continuity at the boundary. Then small holes and objects are removed that would have been introduced in previous steps. The outcome of this process is a symmetry-enforced, and topologically connected structure that adheres to the specified crystallographic constraints.

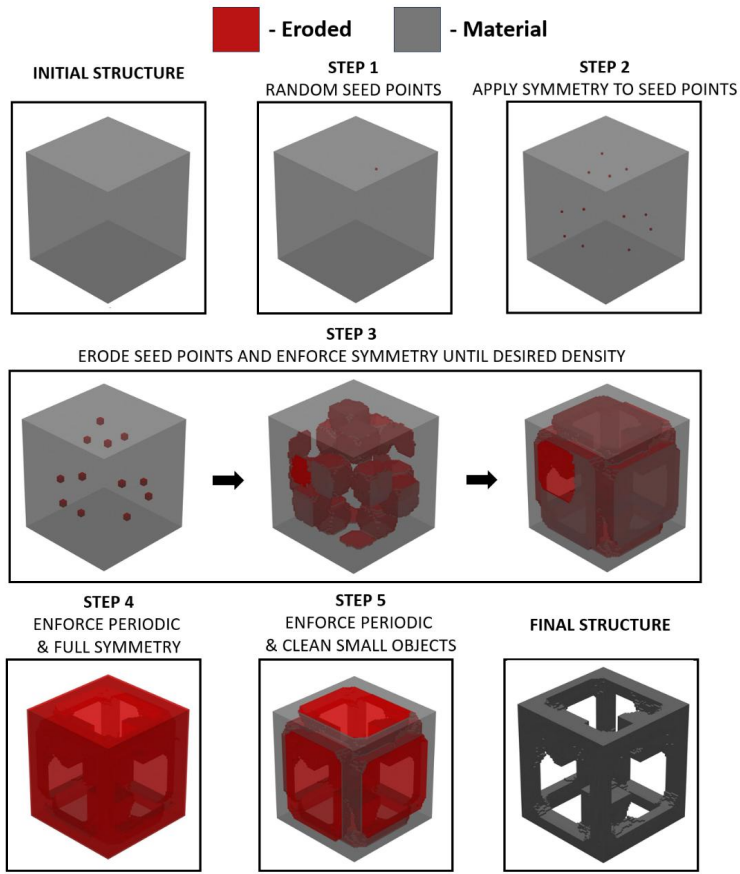

Figure S1: The cubic crystal symmetry-based erosion algorithm for topology generation.

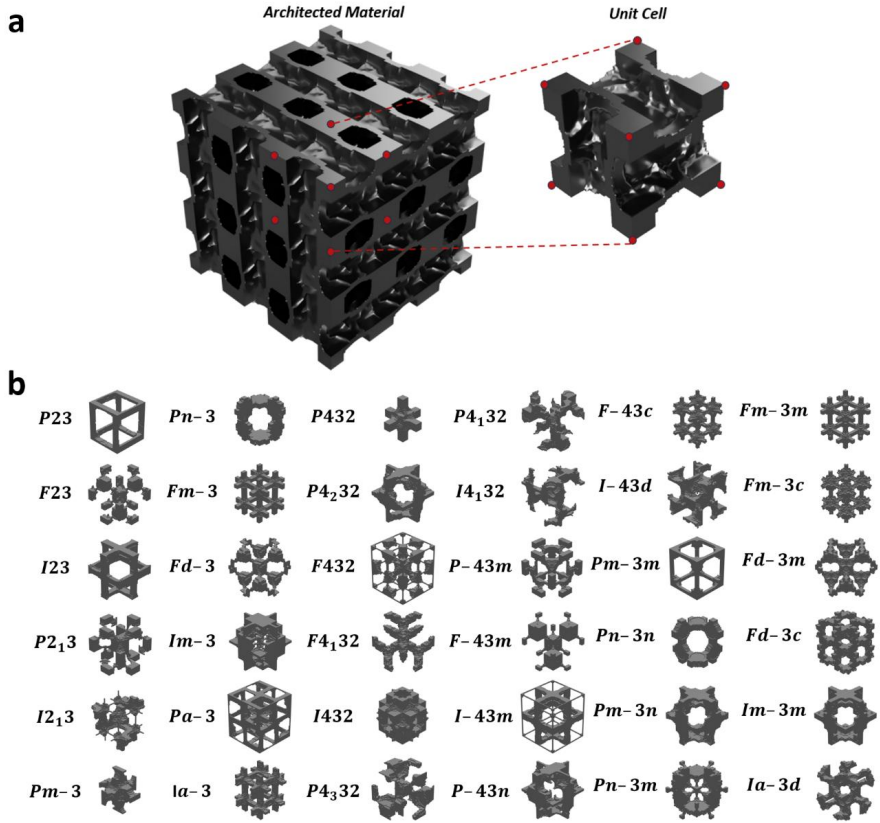

Figure S2: a) Isometric view of an architected material composed of 3x3x3 periodically repeated unit cells. b) Representative samples of cubic unit cells from each of the cubic space groups, ranging from relative densities 0.1 - 0.2.

**Section 2 - Mechanical Properties of Unit Cells in the Dataset**

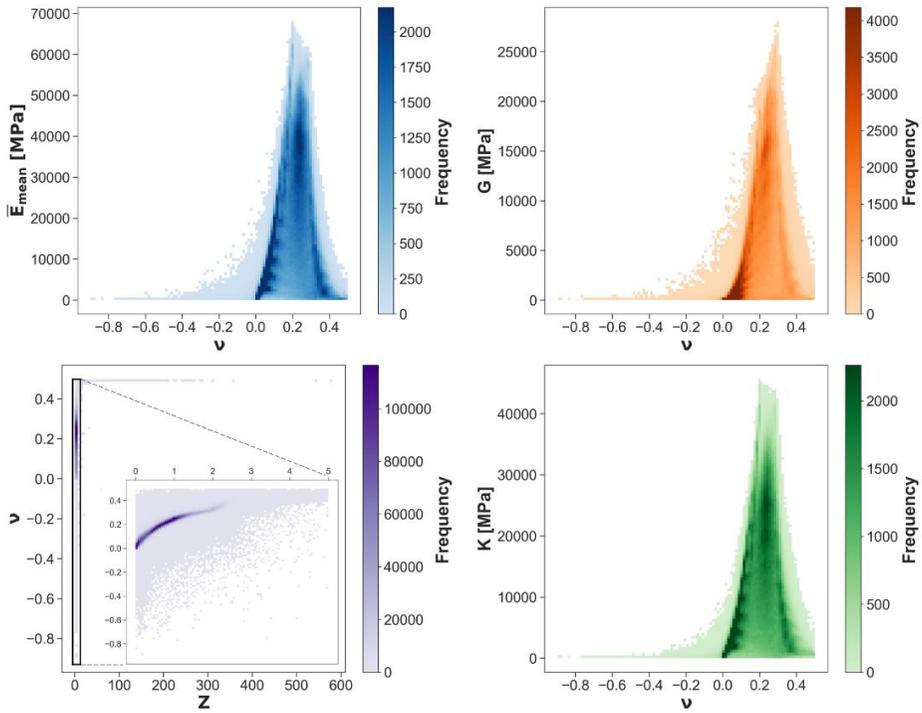

Figure S3: Joint distributions of Poisson's ratio, Zener anisotropy ratio, and effective elastic moduli for the full dataset.

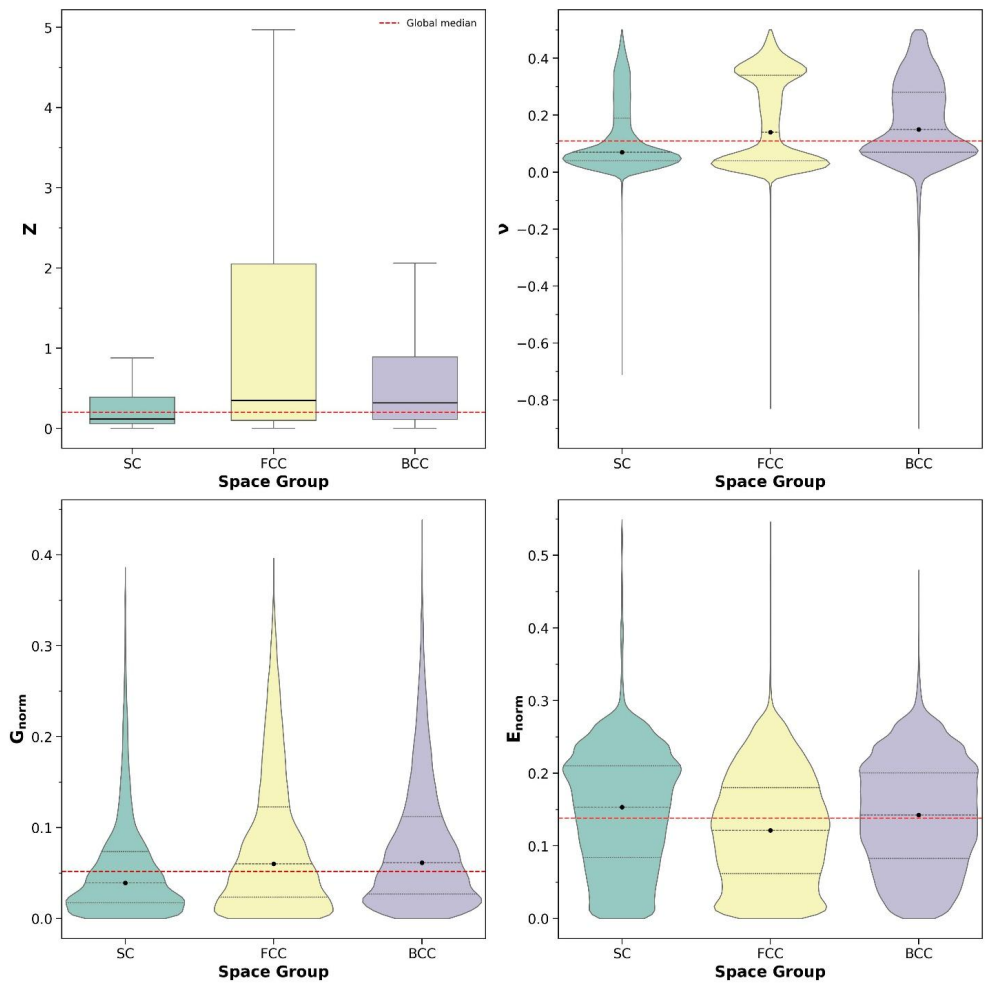

Figure S4: Distributions of elastic moduli and Zener anisotropy across Bravais lattice types within the low relative density regime, balanced by random down sampling ($\rho \in [0.05, 0.2]$)

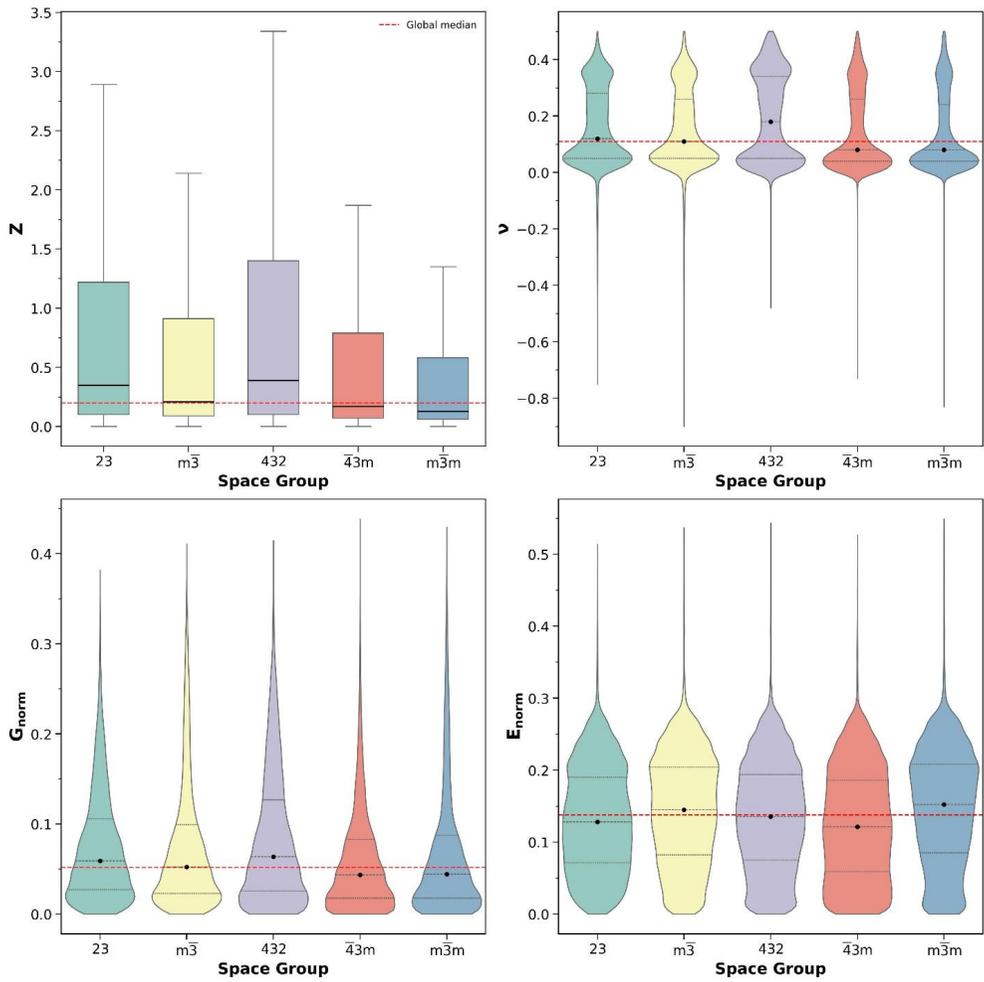

Figure S5: Distributions of elastic moduli and Zener anisotropy across point groups within the low relative density regime, balanced by random down sampling ($\rho \in [0.05, 0.2]$)

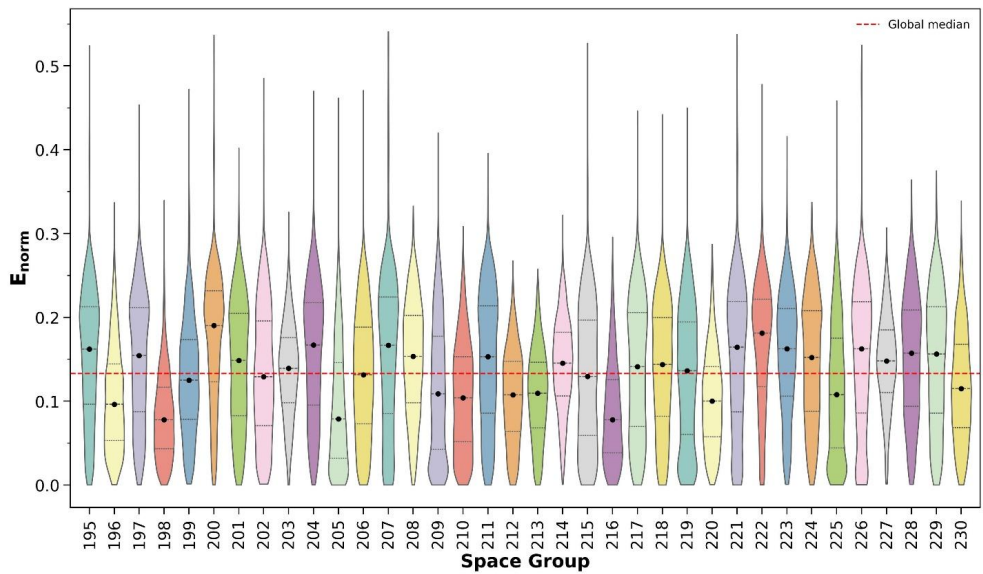

Figure S6: Distributions of $E_{norm}$ across all space groups within the low relative density regime, balanced by random down sampling ($\rho \in [0.05, 0.2]$)

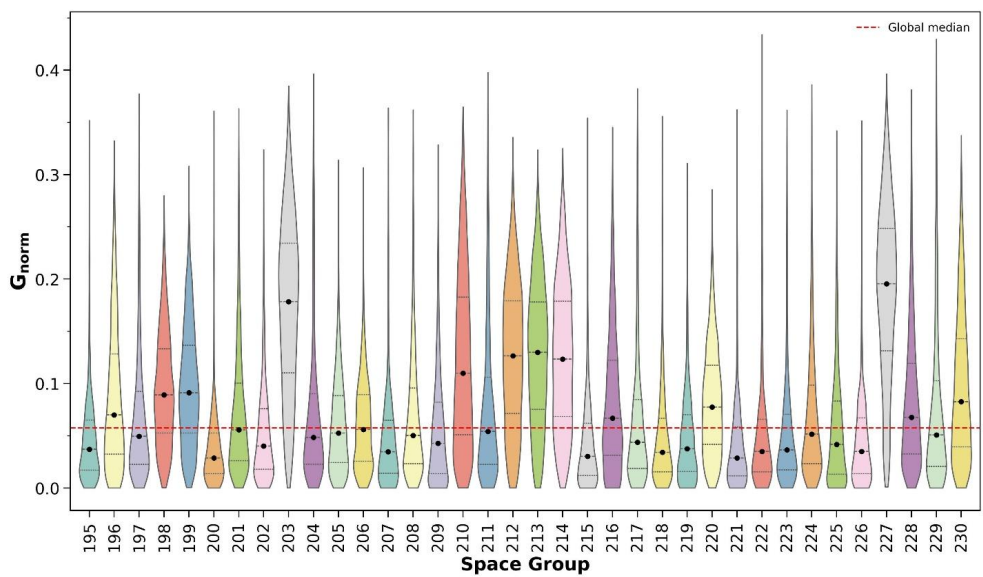

Figure S7: Distributions of $G_{norm}$ across all space groups within the low relative density regime, balanced by random down sampling ($\rho \in [0.05, 0.2]$)

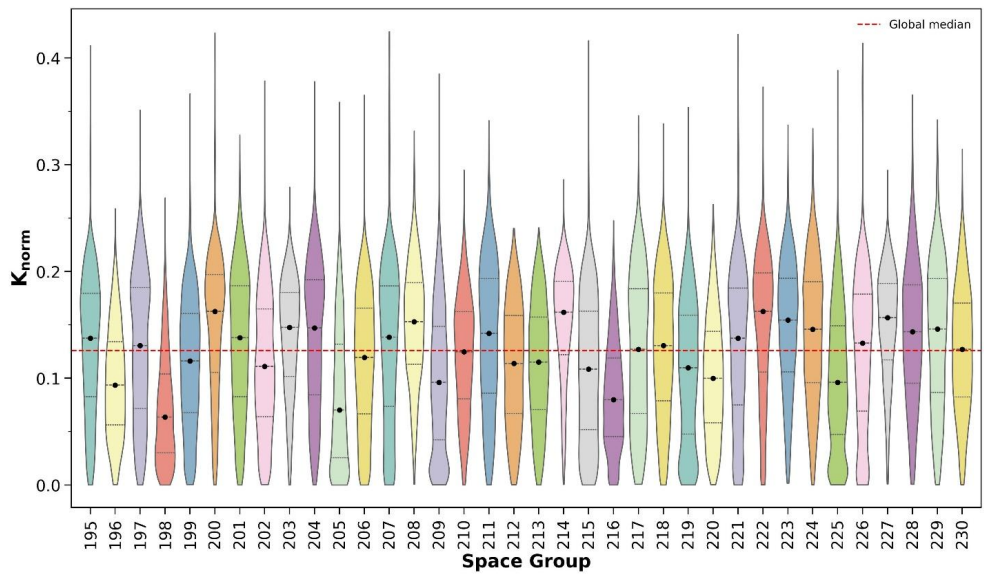

Figure S8: Distributions of $K_{norm}$ across all space groups within the low relative density regime, balanced by random down sampling ($\rho \in [0.05, 0.2]$)

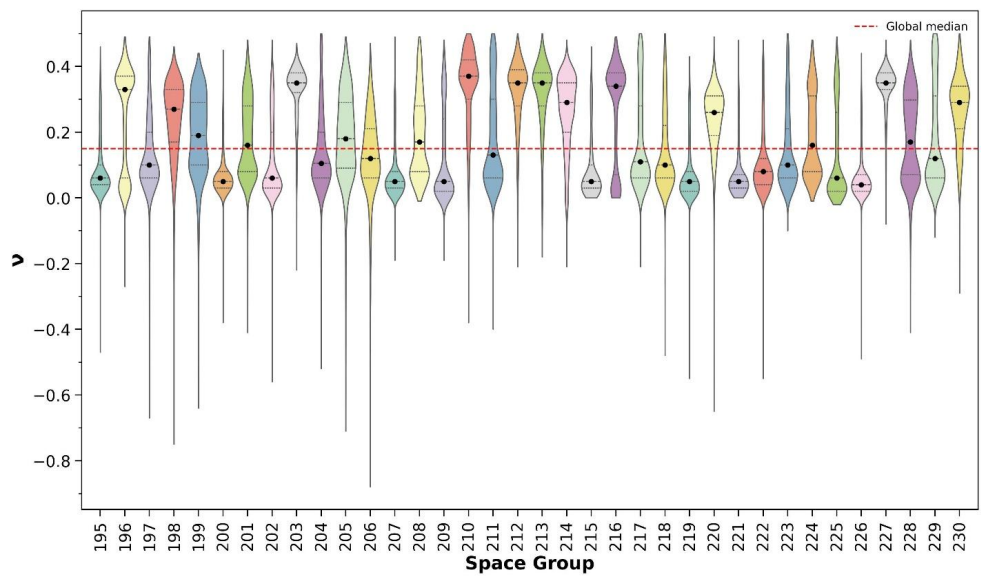

Figure S9: Distributions of $\nu$ across all space groups within the low relative density regime,

balanced by random down sampling (ρ ∈ [0.05, 0.2])

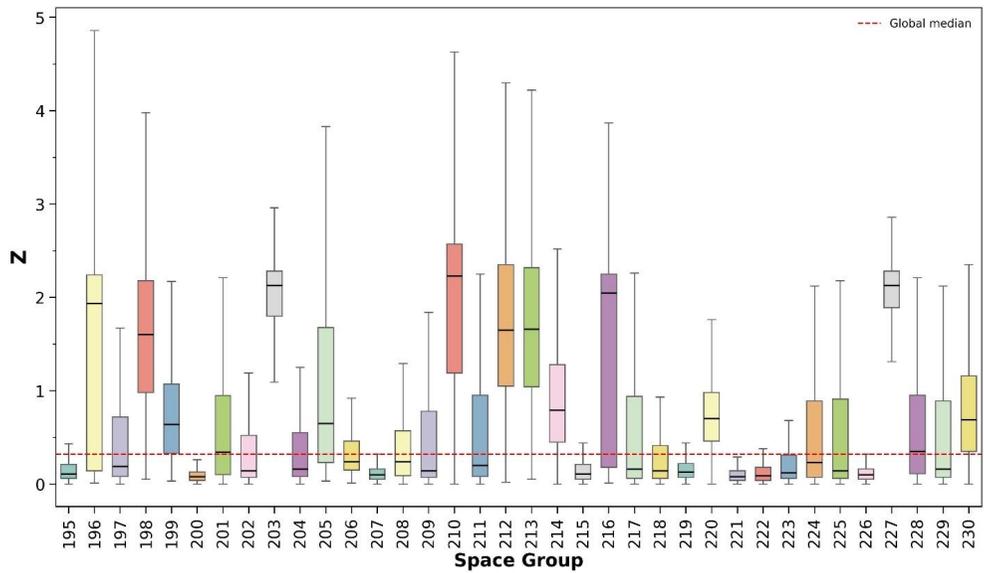

Figure S10: Distributions of $Z$ across all space groups within the low relative density regime, balanced by random down sampling (ρ ∈ [0.05, 0.2])

## Section 3 - Machine Learning

The CNN architecture (Figure S11) is based on the ResNet model [1], utilizing residual connections to enable deep models that efficiently extract hierarchical features while maintaining stable gradient flow. The model consists of multiple residual blocks, with progressively increasing channel depths, composed of stacked 3 × 3 convolutional layers with batch normalization and ReLU activations, with 1 × 1 convolutional skip connections used to match feature dimensions where required.

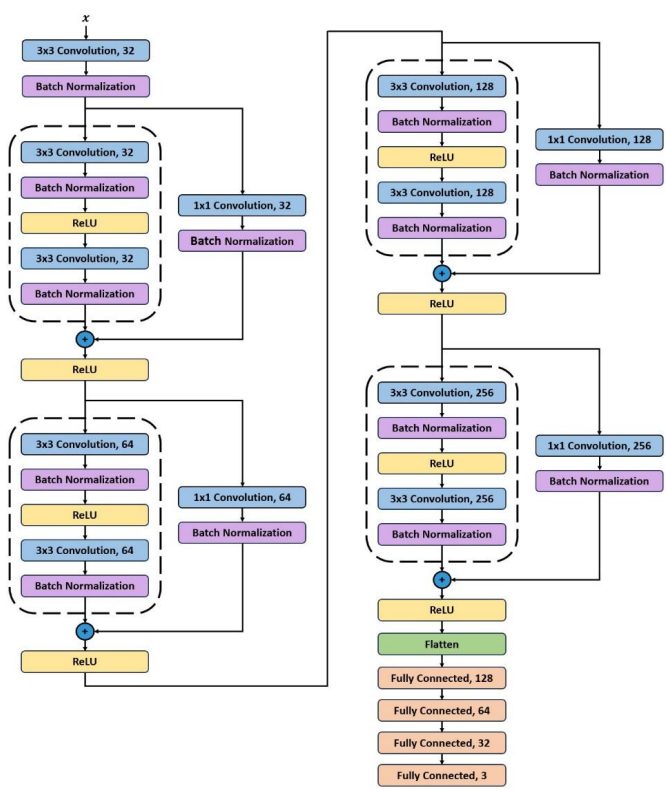

Figure S11: 3D CNN model architecture used to predict the strain energy densities of the voxelated structures.

Both the training and validation losses converge within the first epoch and remain stable and near zero throughout the remainder of training (Figure S12). This rapid convergence is

attributable to the scale of the dataset, as each training epoch consists of approximately 50,000 parameter update steps. The close alignment between training and validation curves indicates that the model generalizes well and is not overfitting.

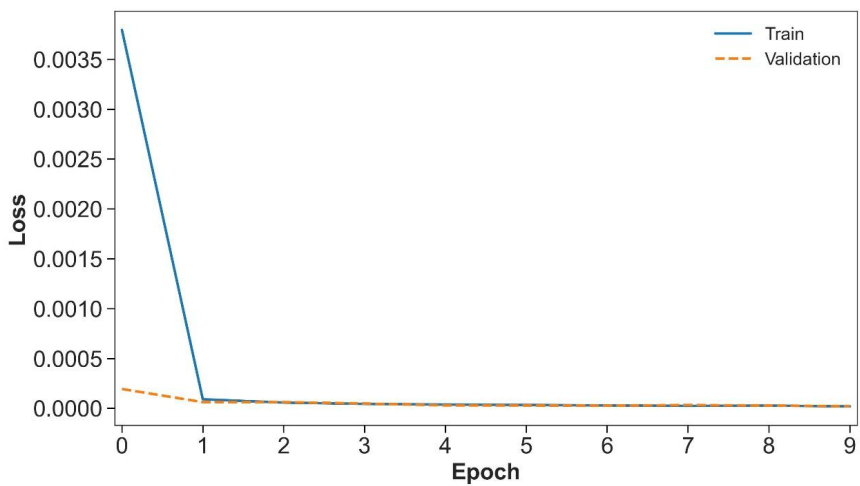

Figure S12: Training and Validation loss for the trained model

To better understand the CNN's capability, we compare performance of the model when trained on lower symmetry structures and evaluated on higher symmetry groups (Figure S13a) with the converse case of training on higher symmetry structures and evaluating on lower symmetry groups (Figure S13b). Although both training cases achieve strong predictive accuracy, models trained exclusively on lower symmetry data clearly outperform those trained solely on higher symmetry data. This demonstrates that lower symmetry structures provide mechanistic information that generalizes well for higher symmetry, but training exclusively on higher symmetry doesn't generalize as well for lower symmetry.

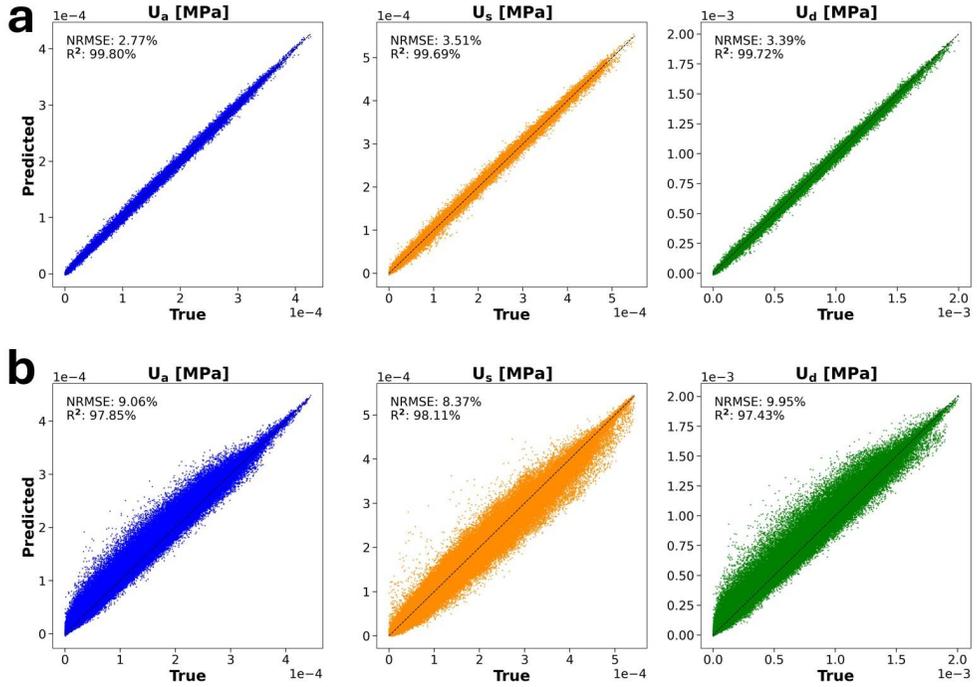

Figure S13: a) Parity plots of the CNN predicting samples from space groups 229 and 230 when trained on 195-200. b) Parity plots of the CNN predicting samples from space groups 195-199 when trained on 221-230.

### Section 4 - Explainability

To gain insight into the model's behavior we use 3D saliency maps to visualize the voxels that are the most relevant for the output of a given input unit cell. To determine the voxels that are the most influential we employed the SmoothGrad technique [2], a gradient-based technique that enhances interpretability by reducing the noise by adding noise to the input. SmoothGrad works by creating multiple copies of the input that is perturbed by adding Gaussian noise, then computing the gradient of the output with respect to each and averaging the resulting gradients to highlight regions most influential to the prediction. The higher the gradients the more influential the voxel is towards the output.

The vanilla gradient saliency map is computed as:

$$S(x) = \left(\frac{\partial f(x)}{\partial x_1}, \frac{\partial f(x)}{\partial x_2}, \ldots, \frac{\partial f(x)}{\partial x_V}\right) \quad (A.1)$$

Where $V$ is the total number of voxels ($V = 65^3$), $x$ is the binary input voxel matrix and $f(x)$ is the scalar prediction from the CNN. Gradients are computed using TensorFlow's gradient tape method.

SmoothGrad then builds on the Vanilla Gradient approach by averaging gradients over $N$ perturbed versions of the input, each generated by adding Gaussian noise, represented as $\mathcal{N}(0, \sigma^2)$. Following the work of Smilkov et al. [2] we use an $N$ value of 50 and set the variance to 0.2.

$$\hat{S}(x) = \frac{1}{N}\sum_{i=1}^{N} S(x + \mathcal{N}(0, \sigma^2))^2 \qquad (A.2)$$

We then apply a mask to the 65x65x65 saliency map to retain only gradients corresponding to voxels where the structure exists, allowing us to isolate the attributions that are physically meaningful. Because the input x is a binary voxel grid with the same dimensions as the saliency map, the masked saliency map is obtained by performing element-wise multiplication between the two.

$$\hat{S}_{masked} = \hat{S} \odot x \qquad (A.3)$$

Additionally, to prevent a small number of voxels with extremely high gradients from making the other meaningful activations appear lower visually, we clipped the gradient values at the 99th percentile value. This step improves interpretability by preserving relative importance while mitigating the influence of outliers. Finally, the gradients are normalized to be between [0, 1].

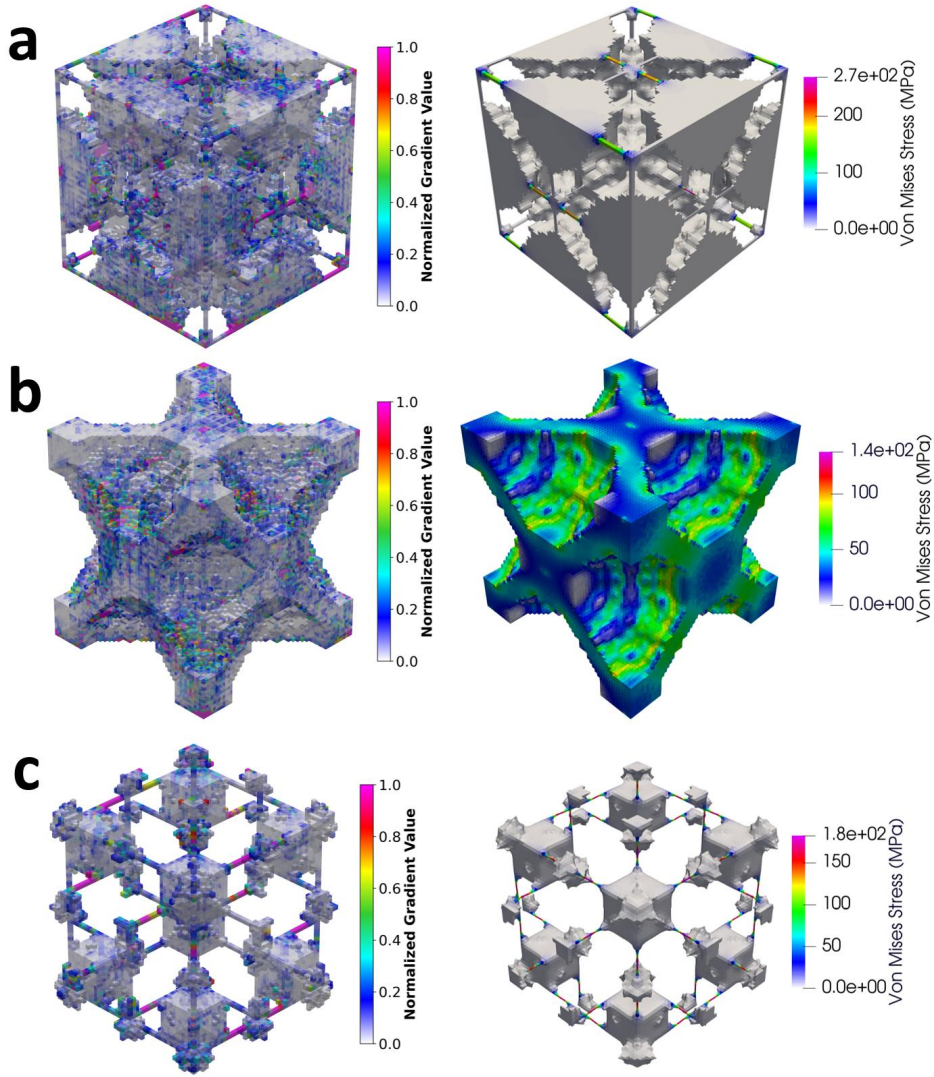

Fig. S14: Saliency maps compared alongside the Von Mises stress contours of extreme structures. a) Highly anisotropic structure. The left side shows the saliency map computed from gradients of the predicted Young's modulus. The right side shows the von Mises stress contour under uniaxial loading. b) Isotropic-optimal structure. The left side shows the saliency map computed from gradients of the predicted Young's modulus. The right side shows the von Mises stress contour under uniaxial loading. c) Pentamode structure. The left side shows the saliency map computed from gradients of the predicted bulk modulus. The right side shows the von Mises stress contour under hydrostatic loading.

## Section 5 - Experimental Validation

The expected weights of the bulk specimens were calculated by multiplying the sample volume with the density of PLA+ (1.23 g/cm³) [3]. For each of the samples printed, the expected mass was estimated as $mass = V_T \bar{\rho} \rho_s$. Where $V_T$ is the total sample volume, $\bar{\rho}$ is the relative density, and $\rho_s$ is the density of the base material.

Table S1: Mass of bulk and lattice samples, all reported in grams

| Sample # | Bulk | Iso-Opt | Simple | OT |
|---|---|---|---|---|
| 1 | 9.24 | 18.65 | 9.19 | 5.83 |
| 2 | 9.27 | 18.64 | 9.21 | 5.82 |
| 3 | 9.25 | 18.65 | 9.21 | 5.82 |
| 4 | 9.36 | 18.65 | 9.20 | 5.83 |
| 5 | 9.35 | 18.64 | 9.20 | 5.83 |
| 6 | 9.34 | - | - | - |
| 7 | 9.26 | - | - | - |
| 8 | 9.29 | - | - | - |
| 9 | 9.40 | - | - | - |
| 10 | 9.25 | - | - | - |
| Average ± std. dev. | 9.30 ± 0.06 | 18.65 ± 0.01 | 9.20 ± 0.01 | 5.83 ± 0.01 |
| Expected Mass | 10.08 | 16.27 | 9.96 | 6.64 |

Table S2: Compression Modulus of bulk and lattice samples, all reported in MPa

| Sample # | Bulk | Iso-Opt | Simple | OT |
|---|---|---|---|---|
| 1 | 1218.41 | 210.92 | 160.37 | 70.78 |
| 2 | 1217.25 | 207.27 | 158.50 | 70.83 |
| 3 | 1239.16 | 208.84 | 158.50 | 70.59 |
| 4 | 1264.69 | 211.14 | 159.40 | 70.80 |
| 5 | 1258.47 | 207.26 | 160.03 | 70.26 |
| 6 | 1249.48 | - | - | - |
| 7 | 1243.71 | - | - | - |
| 8 | 1212.24 | - | - | - |
| 9 | 1263.45 | - | - | - |
| 10 | 1218.33 | - | - | - |
| Average ± std. dev. | 1238.52 ± 20.57 | 209.08 ± 1.89 | 159.36 ± 0.86 | 70.65 ± 0.24 |

Table S3: Relative Stiffness of Lattice Samples

| Sample | Measured Relative Stiffness | Expected Relative Stiffness |
|---|---|---|
| Iso-Opt | 0.17 | 0.30 |
| Simple | 0.13 | 0.14 |
| OT | 0.06 | 0.04 |

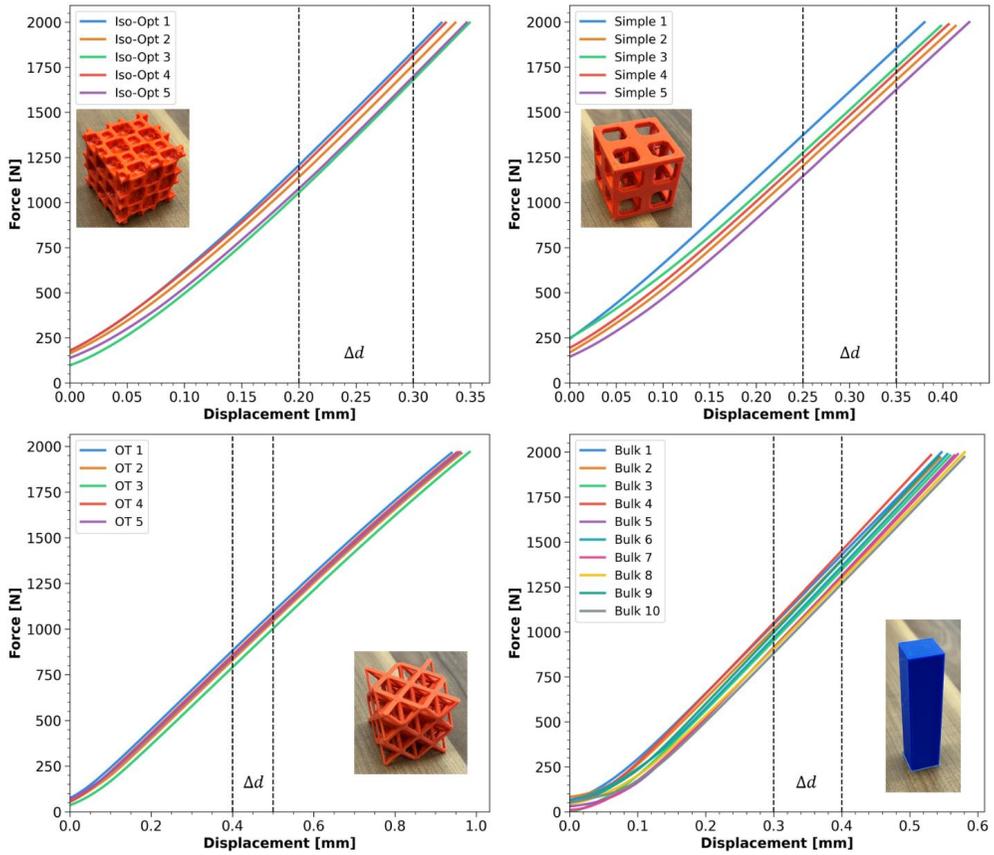

Fig. S15: Force - displacement curves of the tested FFF printed samples. The displacement range used to compute the Young's modulus is denoted by $\Delta d$.

**Table S4**: Sample unit cells of structures from space groups 195–230, with relative densities ranging from 0.1 to 0.5. Shown are the auxiliary, top, right, and front views.

| ID | Space Group | Relative Density | Auxiliary | Top | Right | Front |
|----|-------------|------------------|-----------|-----|-------|-------|
| 1 | 195 | 0.104 | 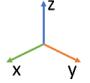 | 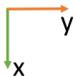 | 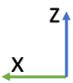 | 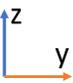 |
| 2 | 195 | 0.202 | 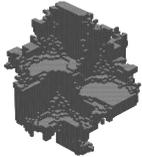 | 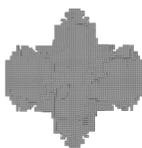 | 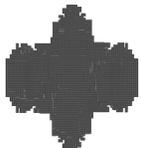 | 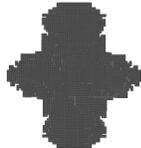 |
| 3 | 195 | 0.300 | 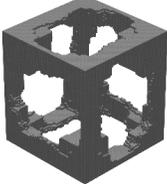 | 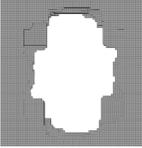 | 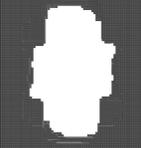 | 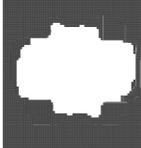 |
| 4 | 195 | 0.396 | 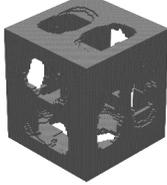 | 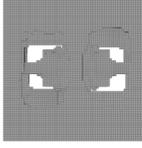 | 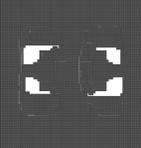 | 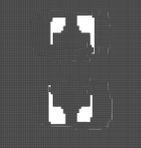 |
| 5 | 195 | 0.499 | 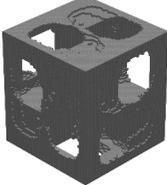 | 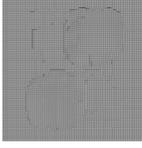 | 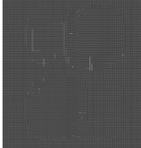 | 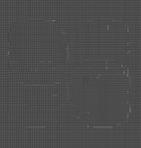 |

| | | | | | | |
|---|---|---|---|---|---|---|
| 6 | 199 | 0.102 | 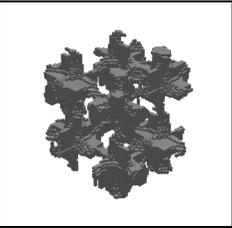 | 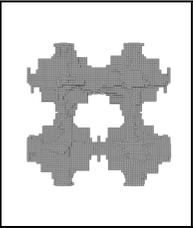 | 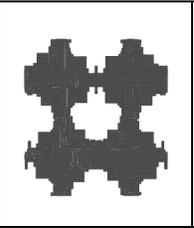 | 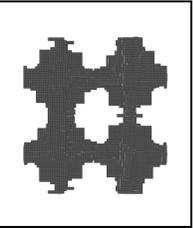 |
| 7 | 199 | 0.196 | 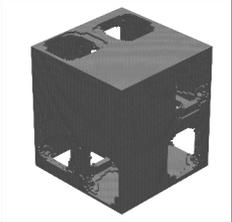 | 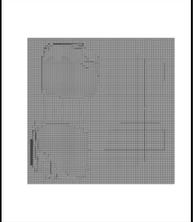 | 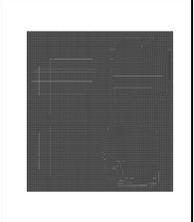 | 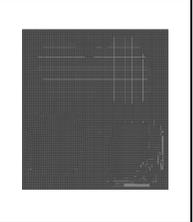 |
| 8 | 199 | 0.300 | 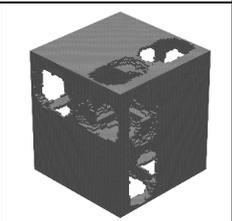 | 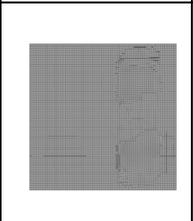 | 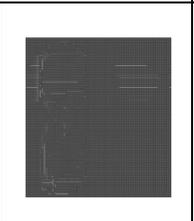 | 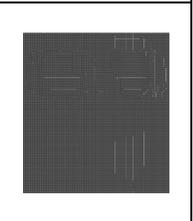 |
| 9 | 199 | 0.397 | 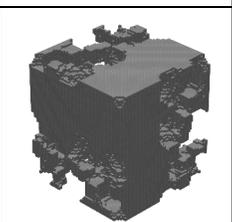 | 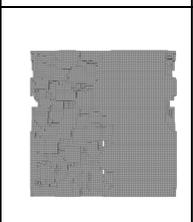 | 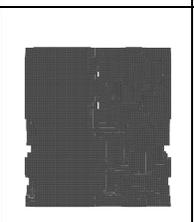 | 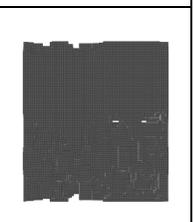 |
| 10 | 199 | 0.497 | 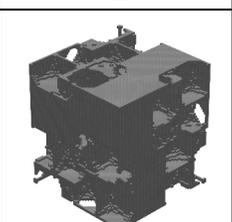 | 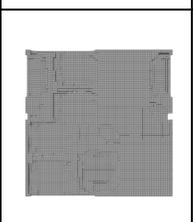 | 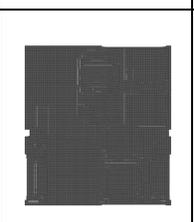 | 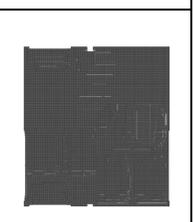 |
| 11 | 200 | 0.102 | 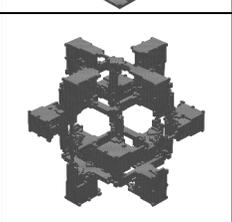 | 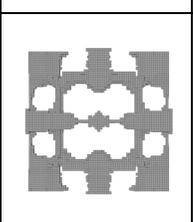 | 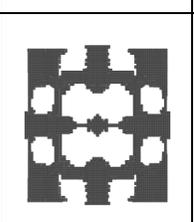 | 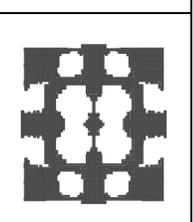 |

| 12 | 200 | 0.197 | 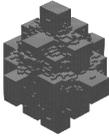 | 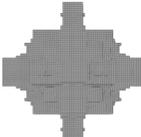 | 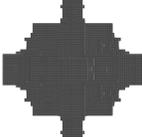 | 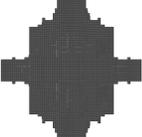 |
|----|-----|-------|---|---|---|---|
| 13 | 200 | 0.295 | 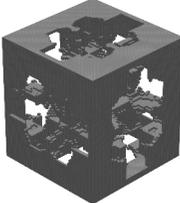 | 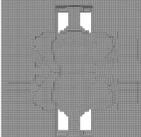 | 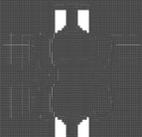 | 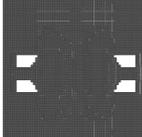 |
| 14 | 200 | 0.400 | 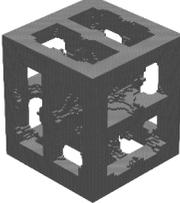 | 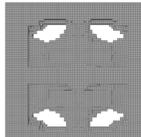 | 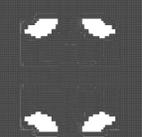 | 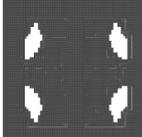 |
| 15 | 200 | 0.497 | 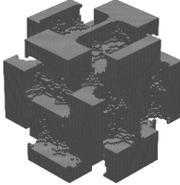 | 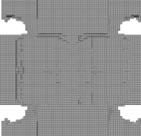 | 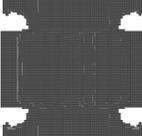 | 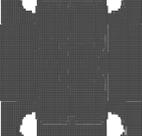 |
| 16 | 206 | 0.098 | 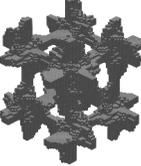 | 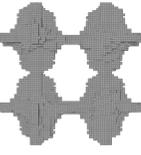 | 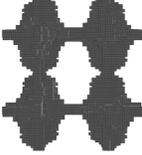 | 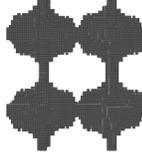 |
| 17 | 206 | 0.198 | 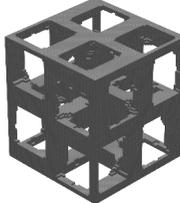 | 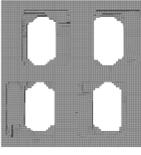 | 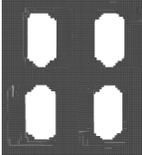 | 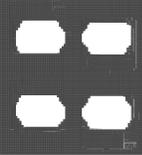 |

| 18 | 206 | 0.296 | 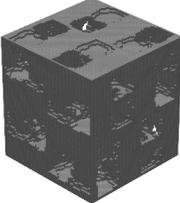 | 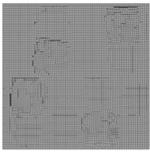 | 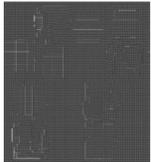 | 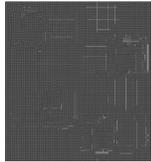 |
|----|-----|-------|---|---|---|---|
| 19 | 206 | 0.404 | 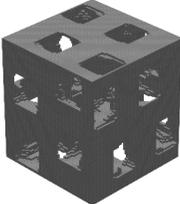 | 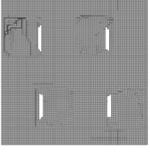 | 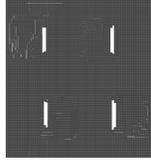 | 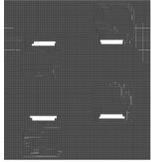 |
| 20 | 206 | 0.499 | 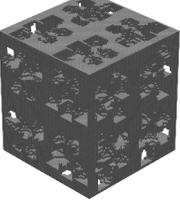 | 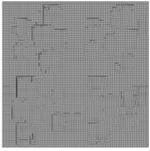 | 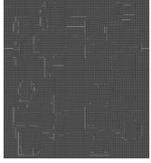 | 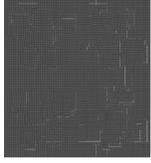 |
| 21 | 207 | 0.095 | 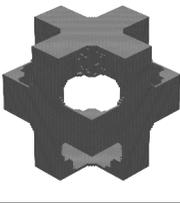 | 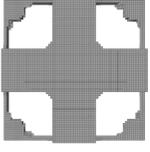 | 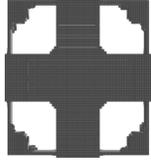 | 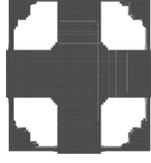 |
| 22 | 207 | 0.200 | 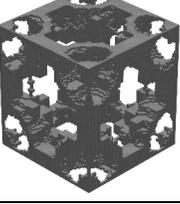 | 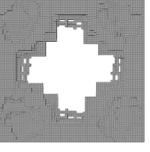 | 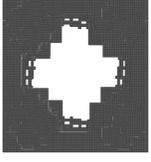 | 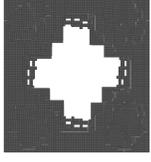 |
| 23 | 207 | 0.297 | 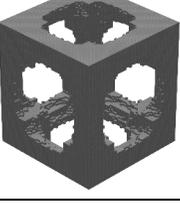 | 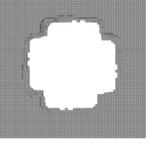 | 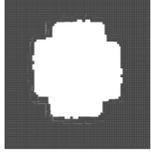 | 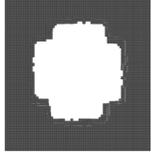 |

| | | | | | | | |
|---|---|---|---|---|---|---|---|
| 24 | 207 | 0.404 | 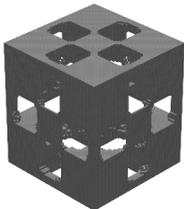 | 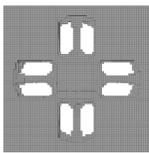 | 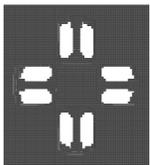 | 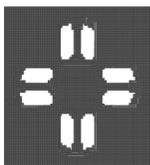 |
| 25 | 207 | 0.499 | 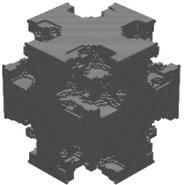 | 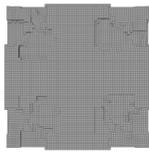 | 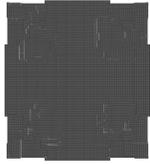 | 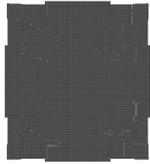 |
| 26 | 214 | 0.103 | 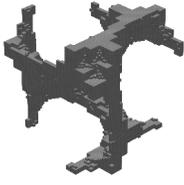 | 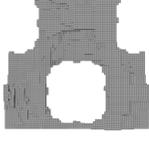 | 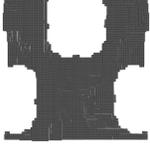 | 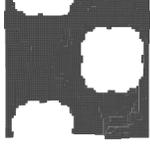 |
| 27 | 214 | 0.197 | 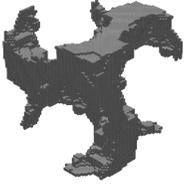 | 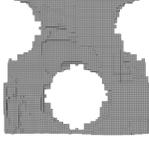 | 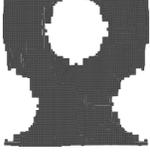 | 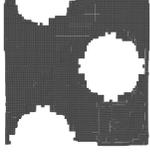 |
| 28 | 214 | 0.304 | 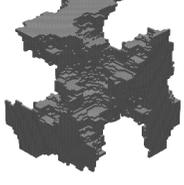 | 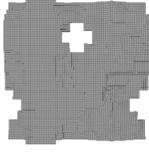 | 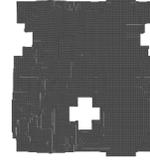 | 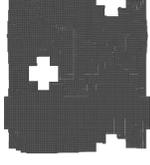 |
| 29 | 214 | 0.404 | 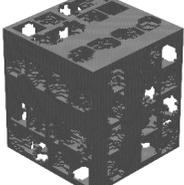 | 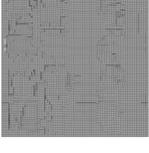 | 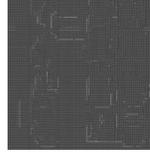 | 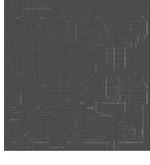 |

| | | | | | | | |
|---|---|---|---|---|---|---|---|
| 30 | 214 | 0.498 | 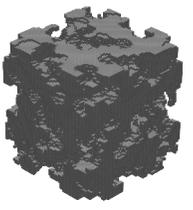 | 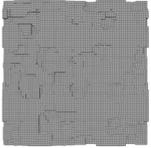 | 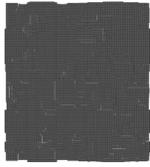 | 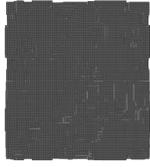 | |
| 31 | 215 | 0.103 | 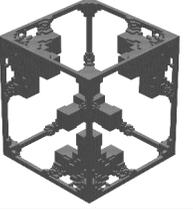 | 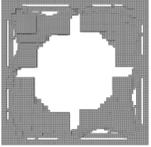 | 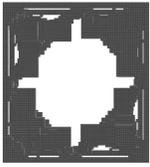 | 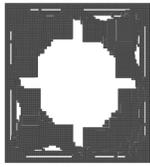 | |
| 32 | 215 | 0.202 | 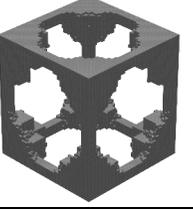 | 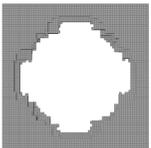 | 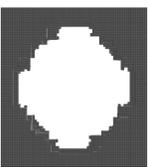 | 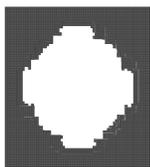 | |
| 33 | 215 | 0.296 | 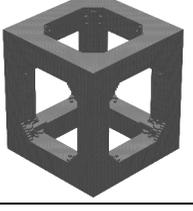 | 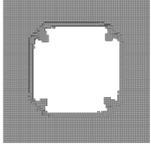 | 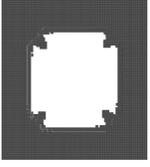 | 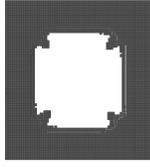 | |
| 34 | 215 | 0.396 | 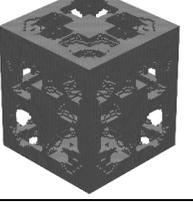 | 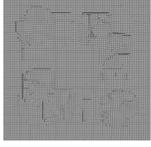 | 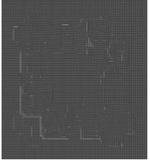 | 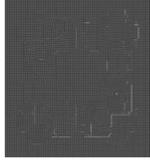 | |
| 35 | 215 | 0.495 | 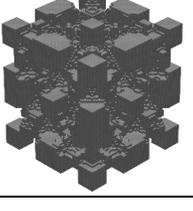 | 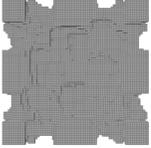 | 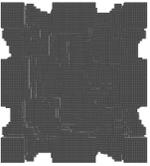 | 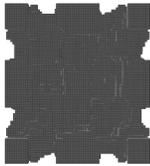 | |

| 36 | 220 | 0.105 | 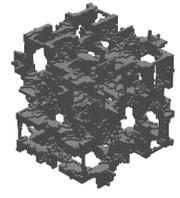 | 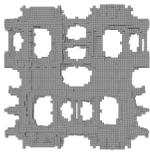 | 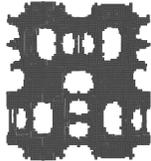 | 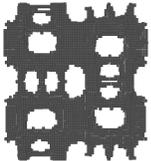 |
|----|-----|-------|---|---|---|---|
| 37 | 220 | 0.197 | 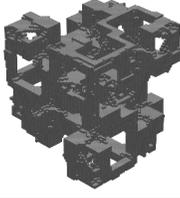 | 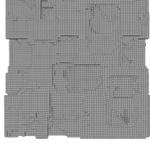 | 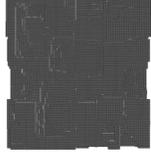 | 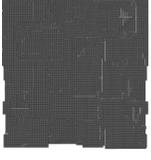 |
| 38 | 220 | 0.295 | 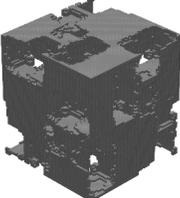 | 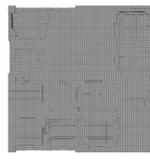 | 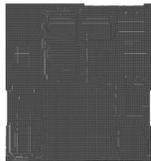 | 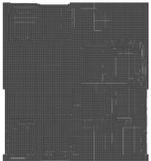 |
| 39 | 220 | 0.397 | 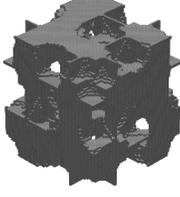 | 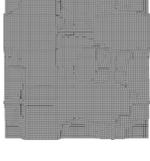 | 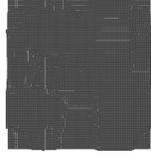 | 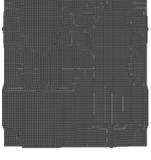 |
| 40 | 220 | 0.497 | 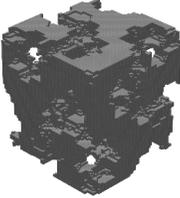 | 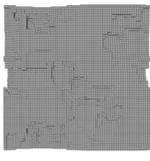 | 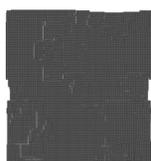 | 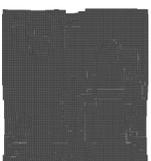 |
| 41 | 221 | 0.103 | 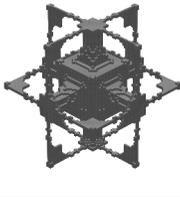 | 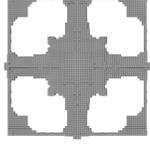 | 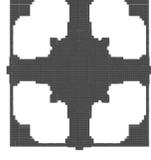 | 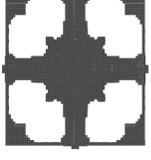 |

| 42 | 221 | 0.201 | 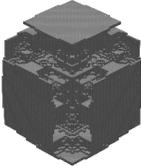 | 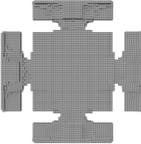 | 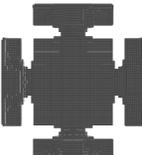 | 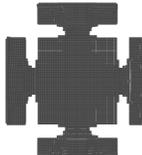 |
|---|---|---|---|---|---|---|
| 43 | 221 | 0.303 | 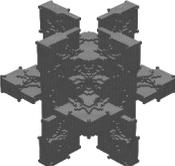 | 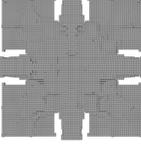 | 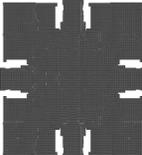 | 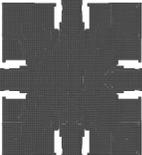 |
| 44 | 221 | 0.403 | 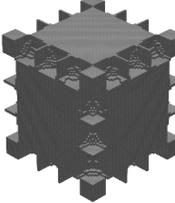 | 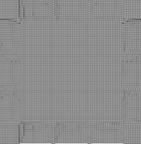 | 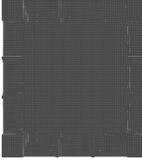 | 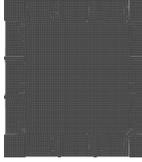 |
| 45 | 221 | 0.499 | 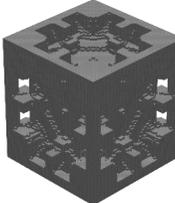 | 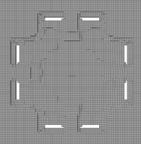 | 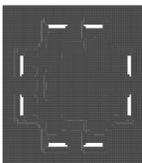 | 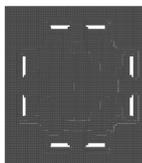 |
| 46 | 230 | 0.104 | 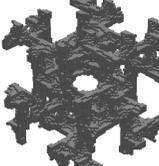 | 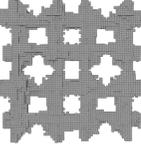 | 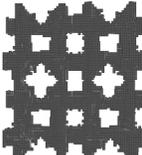 | 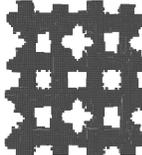 |
| 47 | 230 | 0.202 | 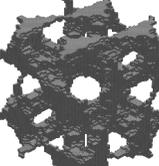 | 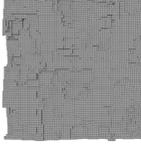 | 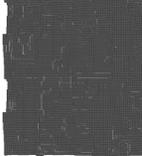 | 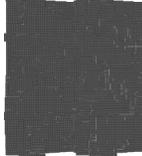 |

| 48 | 230 | 0.295 | 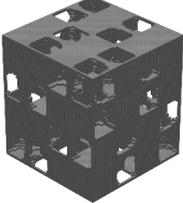 | 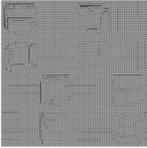 | 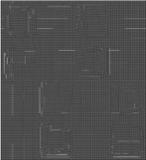 | 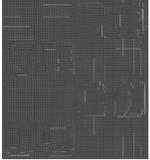 |
| --- | --- | --- | --- | --- | --- | --- |
| 49 | 230 | 0.403 | 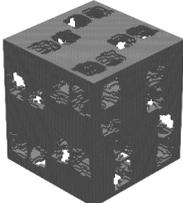 | 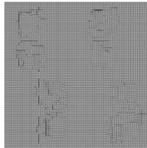 | 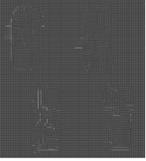 | 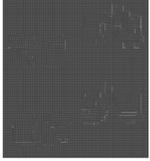 |
| 50 | 230 | 0.500 | 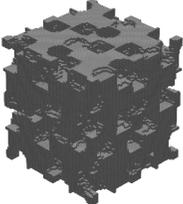 | 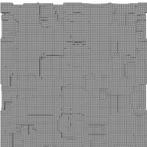 | 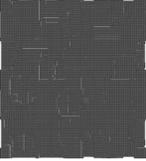 | 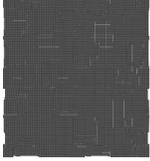 |



% Submissions are not required to reflect the precise reference formatting of the journal (use of italics, bold etc.), however it is important that all key elements of each reference are included.
\newpage
\section {References}
\vspace{-\baselineskip}
\renewcommand{\refname}{}
\bibliography{bibs.bib}

\newpage
\end{document}